\documentclass[paper,11pt]{JHEP}
\pdfoutput=1
\usepackage{graphicx}
\usepackage[centertags]{amsmath}
\usepackage{amsfonts}
\usepackage{amssymb} 
\usepackage{amsthm}
\usepackage{slashed}
\usepackage{enumerate,enumitem}
\usepackage{amsbsy}

\def\({\left(}
\def\){\right)}
\def\[{\left[}
\def\]{\right]}

\def\one{{\rm 1\kern -.9mm l}}                             %

\def\beq{\begin{equation}}
\def\eeq{\end{equation}}
\def\beqa{\begin{eqnarray}}
\def\eeqa{\end{eqnarray}}
\newcommand{\eqa}{\begin{eqnarray}}
\newcommand{\ena}{\end{eqnarray}}

\newcommand{\ho}{{\mathcal{Z}}}
\newcommand\blank[1]{}

\newcommand{\bP}{ {\bf P } }

\newcommand{\balpha}{\alpha\kern -6.7pt\alpha}
\newcommand{\bbalpha}{\alpha\kern -4.95pt\alpha}

\newcommand\en{\end{equation}}
\newcommand\bea{\begin{eqnarray}}
\newcommand\eea{\end{eqnarray}}
\newcommand\nn{\nonumber}

\newcommand{\One}{{\hbox{{\rm 1{\hbox to 1.5pt{\hss\rm1}}}}}}
\renewcommand{\One}{{\mathbb 1}}
\renewcommand{\One}{{\rm 1\!\!1}}

\newcommand{\ba}{\begin{eqnarray}}
\newcommand{\ea}{\end{eqnarray}}

\newcommand{\be}{\begin{equation}}
\newcommand{\ee}{\end{equation}}

\renewcommand{\log}{\ln}

\newcommand{\sig}{\sigma}

\newcommand{\GamOne}{\Gamma}

\setlist[itemize]{leftmargin=*}

\def\XXint#1#2#3{{\setbox0=\hbox{$#1{#2#3}{\int}$ }
\vcenter{\hbox{$#2#3$ }}\kern-.6\wd0}}

\title{12 loops and triple wrapping in ABJM theory from integrability}
\author{ \centerline{Lorenzo Anselmetti$^{1}$, Diego Bombardelli$^{2}$, Andrea Cavagli\`a$^{1}$ and Roberto Tateo$^{1}$ } \\
${}^{1}$\sl\small Dipartimento\ di Fisica
and INFN, Universit\`a di Torino, Via P.\
Giuria 1, 10125 Torino, Italy.
\\
${}^{2}$\sl\small Dipartimento di Fisica e Astronomia and INFN, Universit\`a di Bologna, Via Irnerio 46, 40126 Bologna, Italy. 
\\
\vspace{0.3cm}
\centerline{ \hspace{0.3cm}
\it lorenzo.anselmetti@edu.unito.it, diegobombardelli@gmail.com, cavaglia@to.infn.it, tateo@to.infn.it}
}
\abstract{
Adapting a method  recently proposed  by C. Marboe and D. Volin for ${\cal N}$=4 super-Yang-Mills,
we develop an algorithm for a  systematic weak coupling expansion of the spectrum of anomalous dimensions in the   
$sl(2)$-like sector of planar $\mathcal{N}$=6 super-Chern-Simons.     
The method relies on the Quantum Spectral Curve formulation of the problem and the expansion is written in terms of the interpolating function $h(\lambda)$, with coefficients expressible as combinations of Euler-Zagier sums with alternating  signs. 
We present explicit results up to 12 loops (six nontrivial orders) for various twist-1 and  twist-2 operators, corresponding to triple and double wrapping terms, respectively, which are beyond  the reach of the Asymptotic Bethe Ansatz as well as L\"uscher's corrections.
 The algorithm works for generic states in this sector and in principle can be used to compute arbitrary orders of the weak coupling expansion.
 For   the simplest operator with $L$=1 and   spin $S$=1,  the Pad\'e  extrapolation of the 12-loop result nicely agrees with the available  Thermodynamic Bethe Ansatz data in a relatively wide range of values of the coupling. 
 A \emph{Mathematica}  notebook with a selection of results is attached.
}
\begin{document}
\section{Introduction}
The discovery of integrability in the AdS/CFT   context \cite{Minahan:2002ve} has led
to  important perturbative  and nonperturbative  results for a special selection of  quantum gauge models  
in 2, 3  and 4 space-time dimensions \cite{Beisert:2010jr}. 
The most studied examples are  ${\cal N}$=4 super-Yang-Mills (SYM),  a natural  supersymmetric  generalization of 
Quantum Chromodynamics (QCD),  and  the  ${\cal N}$=6 super-Chern-Simons (SCS) theory in 3d, the so-called  
ABJM model \cite{Aharony:2008ug}.  
Various aspects of these systems  were  scrutinized with the 
aid of integrable model techniques, such as the exact world-sheet and spin-chain S-matrix \cite{Staudacher:2004tk,Arutyunov:2004vx,Beisert:2005tm, Beisert:2006qh,Beisert:2006ez, Janik:2006dc, Arutyunov:2006iu, Arutyunov:2006yd}, the Asymptotic
and the Thermodynamic   Bethe Ans\"atze \cite{Minahan:2002ve, Beisert:2003yb, Beisert:2006ez, Gromov:2009tv,Bombardelli:2009ns, Gromov:2009bc, Arutyunov:2009ur} for   anomalous dimensions of local gauge-invariant operators, the  quark-antiquark
potential \cite{Correa:2012hh,Drukker:2012de} and  the expectation values of Wilson loops in the  strong coupling limit \cite{Alday:2009dv, Toledo:2014koa}.
Three-point functions \cite{Basso:2015zoa, Balitsky:2015tca} and the  generalization, 
beyond the strong coupling limit, of polygonal Wilson 
loops \cite{Basso:2013vsa, Basso:2014hfa, Fioravanti:2015dma} are currently the object  of intense research.    

At present, considering the  results on perturbative \cite{Marboe:2014gma,Marboe:2014sya}, exact \cite{Gromov:2014bva, Alfimov:2014bwa} and
high-accuracy  numerical methods \cite{Gromov:2015wca},
the spectral problem associated  to the study of  anomalous dimensions in
planar ${\cal N}$=4 SYM,  can be considered  virtually solved.
One of the  crucial  final steps  was  recasting the infinite
set of Thermodynamic Bethe Ansatz (TBA)  equations  into a much simpler
matrix nonlinear Riemann-Hilbert problem,  the so-called  Quantum
Spectral Curve (QSC) \cite{Gromov:2013pga,Gromov:2014caa}.
Starting  from the TBA equations of \cite{Bombardelli:2009ns, Gromov:2009bc, Arutyunov:2009ur}, this  simplification  was 
highly nontrivial: it required  a deep   understanding of the
branching properties of the TBA solutions \cite{Cavaglia:2010nm,Cavaglia:2011kd,Balog:2011nm} and  their  
link with generalized T and Q Baxter's functions \cite{Gromov:2010km,Balog:2011nm,Gromov:2011cx}.\\
Finally,  in the paper \cite{Marboe:2014gma}, C. Marboe and D. Volin proposed  an  iterative procedure for the exact 
determination of  perturbative  contributions, which was applied to compute up to 10 loops for many  interesting operators.
These results correspond to the sum of a huge number of  Feynman diagrams and  will probably remain -- for a  very long time --  out of reach of the more standard    approaches to  quantum field theory.  

The ABJM theory in 3d is a second  interesting  example of quantum  gauge
theory which can be considered  to be exactly solvable  in the planar limit. 
An exact description of the spectrum of anomalous dimensions was
obtained by combining information from two-loop perturbation theory \cite{Minahan:2008hf} and on the
strong coupling limit, corresponding to the classical limit
of type IIA superstring theory on $AdS_4 \times  CP^3$ \cite{Stefanski:2008ik, Arutyunov:2008if,Gromov:2008bz}.\\
This led to a conjecture for  the  Asymptotic Bethe Ansatz equations  \cite{Gromov:2008qe},
describing operators with large quantum numbers, and to a proposal for the
vacuum and the simplest excited state TBA equations \cite{Bombardelli:2009xz, Gromov:2009at}. The TBA   formally   encodes  the full information on the 
anomalous dimensions in terms of the dressed coupling constant $h$, the so-called interpolating function which is an essential ingredient of  the integrability approach to this model. As first  noticed in \cite{Gaiotto:2008cg}, 
$h(\lambda)$  is a nontrivial function of the  t'Hooft
coupling $\lambda$ as, for example, it scales differently  with $\lambda$ at weak coupling ($h \sim \lambda$) and at strong coupling  ($h \sim \sqrt{\lambda/2}$). For the shortest unprotected operator, the so-called {\bf 20}, the TBA equations   were solved numerically at intermediate values of $h$ 
 in \cite{LevkovichMaslyuk:2011ty}.\\
However,  considering the parallel  progress made for   ${\cal N}$=4  SYM,  these remarkable achievements  cannot be considered fully satisfactory and the
purpose of \cite{Cavaglia:2013hva}  was to begin a   reduction procedure which ultimately led to the Quantum Spectral Curve formulation of the spectral problem \cite{Cavaglia:2014exa}.
Starting  from the  latter set of equations,   exact  results  on the slope function were obtained in \cite{Gromov:2014eha}, together with an  
important conjecture on the precise dependence of $h$ on the t'Hooft coupling $\lambda$. 

The main  purpose of this article   is  to show that the perturbative scheme described in \cite{Marboe:2014gma}  can be  adapted to the study of the ABJM theory. 
 We will consider a set of $sl(2)$-like states corresponding to single-trace operators of the form \cite{Klose:2010ki}
\beqa
\text{tr}\left[ D_+^S (Y^1 Y_4^{\dagger} )^L \right].
\eeqa
 Rigorously speaking, these states do not form a closed subsector but belong to a wider $OSp(2|2)$ subsector. In terms of the global $OSp(6|4)$ symmetry of the theory, they are characterised by the Dynkin labels $\left[ L + S , S ; L, 0, L \right]$; in particular, the operator { \bf 20 } carries the charges $L=1$, $S=1$. 
 The spectrum of $sl(2)$-like states has been previously studied in \cite{Gromov:2009tv,Beccaria:2009ny,Minahan:2009aq,Minahan:2009wg,Papathanasiou:2009zm,Leoni:2010tb,Beccaria:2010kd,LevkovichMaslyuk:2011ty}, and explicit results were known up to four loops. \\
Although the current  work parallels in many ways the  ${\cal N}$=4 SYM case, the difference in the analytic properties
between the two models makes the study of the ABJM theory nontrivial and still very  challenging, as both  the iterative procedure of \cite{Marboe:2014gma}
 and the associated publicly available \emph{Mathematica} program had  to be revised  step by step. 
 
The algorithm shows that the answer, at a generic loop order, is expressed as a linear combination, with algebraic\footnote{In this paper, we restrict to states characterised by a rational Baxter polynomial. In this case, the coefficients are actually rational.} coefficients,  of alternating Euler-Zagier sums \cite{bailey1994experimental,Borwein:1996yq,Broadhurst:1996az, DataMine}: 
 \beqa\label{eq:sums}
\zeta_{a_1 , \dots , a_k } = \sum_{1  \leq n_1  < \dots <  n_k } \; \prod_{i=1}^k \, \frac{ (\text{sgn}( a_i ) )^{n_i} }{ n_i^{|a_i| } } , \;\;\;\;\; a_j \in \mathbb{Z} \setminus \left\{0\right\}, 
\eeqa
where $a_k \neq 0 $ to avoid divergence.
 In contrast, the results of \cite{leurent2013multiple, Marboe:2014gma} show that anomalous dimensions in the $sl(2)$ sector of $\mathcal{N}$=4 SYM can contain only Euler-Zagier sums with all signs strictly positive, or Multiple Zeta Values (MZV). \\
  Examples of results can be found in Appendix \ref{app:results}.
  We see the appearance of the simplest alternating sum, $\zeta_{-1} = -\log(2)$,  at 6 loops  for twist-1 operators; irreducible multiple sums such as e.g.\footnote{ As a historical remark, this alternating double sum first appeared in the computation of the anomalous magnetic moment of the electron at three loops in QED \cite{Laporta:1996mq}.} $\zeta_{-1,-3}$ start to appear from 8 loops, corresponding to double wrapping. \\
  We point out that the weak coupling expansion of the exact form of $h(\lambda)$  conjectured in \cite{Gromov:2014eha}   yields only $\zeta_{2n} \propto \pi^{2 n}$ terms (see equation (\ref{eq:hlambda})); the general transcendental structure of the results appears therefore very similar when they are expressed in terms of the t'Hooft coupling. Nonetheless, it is certainly worth studying whether this rearrangement of terms can reveal some indirect evidence in support of the conjecture of \cite{Gromov:2014eha}. We leave this for future investigations. \\
   We performed explicit calculations up to 12 loops 
 for the simplest twist-1 and twist-2 operators, however the procedure is valid for any values of the twist and spin, and can be pushed in principle to an arbitrary order. 
 
  Let us collect some comments and observations.   We find that the anomalous dimensions do not contain terms of transcendentality one. 
 It would be interesting to see whether there are further selection rules for the answers. 
 In the case of $\mathcal{N}$=4 SYM, it was observed in \cite{leurent2013multiple} and  \cite{Marboe:2014gma} that, up to 10 loops, only a subclass of the possible combinations of MZVs appear, related to single-valued polylogarithms \cite{2013arXiv1302.6445S, 2013arXiv1309.5309B}, and it is thus natural to wonder whether the same happens for ABJM (either in terms of $h$ or of $\lambda$). 
\\
 Another preliminary observation concerns the  \emph{shift symmetry}: it was noticed in  \cite{Beccaria:2009ny} that the anomalous dimensions for the operators $(L, S)=(1, 2 n)$ and $(L,S)=(1, 2 n-1)$, $n \in \mathbb{N}^{+}$ are the same, at the level of the ABA, up to four loops. Our results suggest that this may be the manifestation of an exact symmetry, holding for the maximum and next-to-maximum transcendentality part\footnote{ An analogous phenomenon may occur for an asymptotic degeneracy of $\mathcal{N}=4$ SYM considered in \cite{Arutyunov:2011uz, Sfondrini:2011rr}. The lifting of the degeneracy due to wrapping corrections was calculated in \cite{Sfondrini:2011rr}, and appears not to affect the highest transcendentality part of the result. } of the full anomalous dimension, including wrapping corrections. So far we checked this explicitly up to 12 loops for $1 \leq S \leq 6$.\\
  Finally, the analysis of the highest transcendentality part of the anomalous dimensions reveals a significant difference between ABJM and $\mathcal{N}$=4 SYM. In the latter case, such terms are completely determined by the ABA and single-wrapping corrections. This property makes it possible to compute this part of the answer as an exact all-loop expression \cite{leurent2013multiple}. 
  In the ABJM case, we discover that the anomalous dimensions are also affected, at all levels of transcendentality, by double and perhaps even higher wrapping corrections\footnote{We are currently unable to disentangle double from higher orders of wrapping. One could try to tackle this problem using the generalised L\"uscher  approach described in \cite{Bombardelli:2013yka}.}. 
  The difference between the two theories in this respect reminds us of the situation with the slope function \cite{Basso:2011rs}, which is a  purely ABA quantity in $\mathcal{N}$=4 SYM but is affected by wrapping corrections in the case of ABJM theory \cite{Beccaria:2012qd,Gromov:2014eha}. 
  It would be nice to see how the computation of \cite{leurent2013multiple} could be generalised in ABJM.

The rest of the paper is organized as follows. In Section \ref{sec:QSC}, we review the QSC construction of \cite{Cavaglia:2014exa} and setup some useful notation. In Section \ref{sec:algo}, we describe the logic of the iterative algorithm.
In Section \ref{sec:results}, we present some checks  of the results, against L\"uscher corrections and by comparison with the TBA data of \cite{LevkovichMaslyuk:2011ty}. Section 
\ref{sec:conclusions} contains the conclusions and some comments on future directions of research. In Appendix \ref{app:results} we list the results for a selection of operators. Appendices \ref{app:inhomo} and \ref{app:functions} contain technical details completing our description of the algorithm. 

\section{The Quantum Spectral Curve equations}\label{sec:QSC}

In this Section, we will recall the QSC formulation introduced in \cite{Cavaglia:2014exa}, valid non-perturbatively for finite values of the coupling $h$, which will be the basis of the algorithm presented in Section \ref{sec:algo}. 

The unknown functions appearing in the problem are defined on the complex domain of the rapidity $u$. At finite $h$, they in general exhibit an infinite  ladder of square-root branch cuts, which are described more precisely below. 
We shall denote by ${\tilde f }$ the second sheet evaluation of a function $f$, obtained by crossing the cut on the real axis, and by $f^{[n]}$ the shifted value  $f(u + i \frac{n}{2} )$ obtained by avoiding all cuts.  

 The functions $\left\{ \bP_i \right\}_{i=0}^4$ are particularly simple since,  on the principal Riemann sheet, they have a single branch cut running along $ u \in (-2 h , 2 h )$. They satisfy the constraint
\beqa\label{eq:quad}
( \bP_0 )^2 &=& 1 - \bP_1 \bP_4 + \bP_2 \bP_3 .
\label{P0}
\eeqa
The functions $\left\{ \nu_i \right\}_{i=1}^4$ instead display  infinitely many cuts for $u \in (-2 h , 2 h ) + i n $,  $n \in \mathbb{Z}$, but have a simple periodicity/anti-periodicity property
\beqa\label{eq:periodic}
{ \tilde \nu }_i = \sigma_i \nu_i^{[2]} = \sigma \, \nu_i^{[2]}, 
\eeqa
where $\sigma = \pm 1 $. The necessity to include anti-periodic solutions (with $\sigma = -1$) was not discussed in \cite{Cavaglia:2014exa}, where the QSC equations were first proposed. Notice however that the introduction of this sign is fully consistent with the construction of \cite{Cavaglia:2014exa}: in fact, the $\nu_a$ functions were introduced to parametrize quadratically a periodic matrix $\mu_{ab}$  which was proved to satisfy $ { \tilde \mu}_{ab} = \mu_{ab}^{[2]}$. It will appear very clearly from the weak coupling analysis that, in order to capture all states, one needs to allow both signs when taking the square root of this relation. The value of $\sigma$ in (\ref{eq:periodic}) depends on the operator under consideration. For instance, for $L=1$ one has $\sigma = (-1)^S$; more generally, $\sigma$ depends on the mode numbers of the state and can be defined in terms of the associated Baxter polynomial, as will be discussed in Section \ref{sec:algo}. 

 The complete analytic structure is described by a set of Riemann-Hilbert type relations
\beqa \label{eq:nut}
{ \tilde \nu}_a &=& - \bP_{ab} \, \chi^{bc} \nu_c  ,\\
{ \widetilde {\bP} }_{ab} - { \bf P }_{ab} &=& \nu_a { \tilde \nu }_b - \nu_b { \tilde \nu }_a , \label{eq:Ptil} 
\eeqa
where 
\beqa
\bP_{ab} = \left( \begin{array}{cccc} 0 & - \bP_1 & - \bP_2 & - \bP_0 \\  \bP_1 & 0 & - \bP_0 & - \bP_3 \\ \bP_2 & \bP_0 & 0 & - \bP_4 \\ \bP_0 & \bP_3 & \bP_4 & 0 \end{array} \right) , \;\;\;\; \chi^{ab} = \left( \begin{array}{cccc} 0 & 0 & 0& - 1 \\  0 & 0 & 1 & 0 \\ 0 & -1 & 0 & 0 \\ 1 &  0& 0 & 0 \end{array} \right) .
\eeqa
In addition to the previous constraints, we impose some regularity requirements on the solutions: the  $\bP$ and the $\nu$ functions should have no poles on any sheet, and should remain bounded at the branch points (locally they behave like $\sqrt{ u \pm 2 h }$).

The global quantum numbers, $L \in \mathbb{N}^+$ (twist), $S \in \mathbb{N}^+$ (spin) and $\Delta$ (conformal dimension) for the states described in this paper are encoded in the asymptotic behaviour of the solution at large $u$, 
through the relations 
\beqa
\bP_a &\simeq& ( A_1 u^{-L}, \,A_2 u^{-L-1}, \,A_3 u^{+L+1}, \,A_4 u^{+L}, \, A_0 u^0) , \label{eq:Pasy}\\
A_1 A_4 &=&-\frac{(-\Delta +L-S) (-\Delta +L+S-1) (\Delta +L-S+1) (\Delta +L+S)}{L^2 (2
   L+1)} \label{eq:A1A4},\\
A_2 A_3 &=&-\frac{(-\Delta +L-S+1) (-\Delta +L+S) (\Delta +L-S+2) (\Delta +L+S+1)}{(L+1)^2 (2 L+1)} \label{eq:A2A3} ,
\eeqa
while the four components of $\nu$ are required to have the leading power-like asymptotics
\beq 
\nu_a \sim \left( u^{\Delta - L} ,u^{\Delta+1}, u^{\Delta} , u^{\Delta+L+1}  \right).
\eeq
 The anomalous dimension $\gamma \sim \text{O}(h^2)$ is defined as in $\Delta = L+S + \gamma$. \\
 For the following, it is important to underline some further algebraic consequences of the equations presented  above. First, using (\ref{eq:nut}) and (\ref{P0}), one finds
\beq \label{eq:nutt}
{ \nu}_a = - \bP_{ab} \, \chi^{bc} \tilde\nu_c ,
\eeq
 and, subtracting (\ref{eq:nutt}), divided by $\bP_{12}$, to (\ref{eq:nut}), shifted by $+i$ and divided by $\bP_{12}^{[2]}$, we obtain the following matrix TQ Baxter equation:
\beq\label{eq:TQ}
{\bf T}_{a}^{b}(u) {\bf Q}_b(u) =  -\left(\bP_{12}^{[1]}(u)\right)^{-1}  {\bf Q}_a^{[2]}(u) +  \left(\bP_{12}^{[-1]}(u) \right)^{-1}  {\bf Q}_a^{[-2]}(u), 
\eeq
with
\beq
{\bf Q}_a(u) \propto \nu_a^{[1]}(u), \,\,\,  {\bf T}_{a}^{b}(u) = -\sigma \,\left(\frac{\bP_{ak}^{[-1]}(u)}{\bP_{12}^{[-1]}(u)}-\frac{\bP_{ak}^{[1]}(u)}{\bP_{12}^{[1]}(u)}\right)\chi^{kb} .
\eeq
\section{Iteration scheme}\label{sec:algo}
In this Section we describe the iterative procedure for computing the weak coupling expansion of  anomalous dimensions, 
\beqa\label{eq:weakexpgamma}
\gamma(h) =  \sum_{n=0}^{\infty} h^{2 + 2 n} \,\gamma_{2n+2} ,
\eeqa
 where, in contrast with the $\mathcal{N}$=4 SYM setup, each term in (\ref{eq:weakexpgamma}) corresponds to \emph{two} loop orders in ABJM theory (odd loops  are vanishing in this model). 

 The procedure takes as input the integer quantum numbers $L$ and $S$. It should be noted that there are in general different states associated to a given  choice of these charges. Each state is expected to be associated unambiguously to a solution of the 2-loop Bethe Ansatz \cite{Minahan:2002ve}, and is thus identified by the Baxter polynomial $Q(u)$. As will be explained below, the latter is selected at the first iteration of the procedure, from the solutions of the 2-loop Baxter equation (\ref{eq:baxter0}).

\subsection{Analytic assumptions}

The weak coupling expansion is based, as in \cite{Marboe:2014gma}, on a specific ansatz for the form of the $\bP$ functions at finite values of $h$. It is enough to specify the ansatz for $\left\{ \bP \right\}_{i=0}^3$, since we will always consider $\bP_4$ to be defined by (\ref{eq:quad}).  
 Without loss of generality, we shall set $A_1 = 1$ and $A_2 = h^2$. This is possible thanks to the symmetries of the QSC equations, analogous to the H-symmetry described in \cite{Gromov:2013pga} (see Appendix \ref{app:symmetry} for more details). The choice of scaling for $A_2$ is important, and comes from the observation that, 
 due to the classical value of the conformal dimensions,
 \beqa
 \Delta = L + S + \text{O}(h^2), 
 \eeqa
 the combination $A_2 A_3$ in (\ref{eq:A2A3}) is $ \text{O}( h^2)$ at $h \sim 0$. 
 
 Introducing the Zhukovsky variable 
\beq\label{eq:x}
x(u)=\frac{u+\sqrt{u^{2}-4h^{2}}}{2h},
\eeq
 we can always assume that, on the first  Riemann sheet characterised by $|x(u)| > 1$, the $\bP$'s can be expanded as 
\beqa
\bP_1 &=& (x h )^{-L} \left( 1 + \sum_{k=1}^{\infty} c_{1, k}(h) \; \frac{h^k}{x^k} \right), \label{eq:ansatz1}\\
\bP_2 &=& (x h )^{-L} \left( \frac{h}{x} + \sum_{k=2}^{\infty} c_{2, k}(h) \; \frac{h^k}{x^k} \right), \label{eq:ansatz2}\\
\bP_0 &=& (x h )^{-L} \left(  M_{L}(u) + \sum_{k=1}^{\infty} c_{0, k}(h) \; \frac{h^k}{x^k} \right), \label{eq:ansatz3}\\
\bP_3 &=& (x h )^{-L} \left(  K_{2 L + 1}(u) + \sum_{k=1}^{\infty} c_{3, k}(h) \; \frac{h^k}{x^k}  \right)\label{eq:ansatz5}.
\eeqa
 Above, $M_{L}(u)$ and $K_{2 L + 1}(u)$ are polynomials in $u$ of degree $L$ and $2L+1$ respectively, with coefficients depending on $h$,
 \beqa\label{eq:polyMK}
 M_L(u) = A_0(h) \, u^L  + \sum_{j= 0}^{L-1}  m_{j}(h) u^{j} , \;\;\;\;\; K_{2 L + 1}(u) = A_3(h) \, u^{2 L+1} \, + \sum_{j= 0}^{2 L} k_{j}(h) \, u^j ,
 \eeqa
where   $A_0(h)$ and $A_3(h)$ are the same as in (\ref{eq:Pasy}), and we are simply introducing the dependence on $h$ for clarity. 
It is by determining these coefficients order by order in $h$ that one is able to compute the anomalous dimensions using (\ref{eq:A1A4}),(\ref{eq:A2A3}). \\
 We shall make a crucial assumption on  the behaviour of the polynomials, as well as all the coefficients of the expansion at weak coupling:
 \beqa
M_L(u) , \; K_{2 L + 1}(u) , \;  \text c_{i, j}(h) \sim \text{O}( h^0 ) , \;\;\; \text{ for } h \sim 0 . \label{eq:polyns}
 \eeqa 
 This is the same scaling that turned out to be appropriate for the $sl(2)$ sector of $\mathcal{N}$=4 SYM theory described in \cite{Marboe:2014gma}. 
 
To setup an iterative algorithm, we consider the perturbative expansions of the $\bP$ and $\nu$ functions 
for $h\sim 0$, keeping fixed the value of the rapidity $u$. 
These expansions are called \emph{normal scaling} expansions \cite{Marboe:2014gma} and read more explicitly
\beqa
{\bf  p}_{i, \text{ns}}(u) &=& \sum_{j=0}^{\infty} h^{2 j} \, {\bf  p}_{i, \text{ns}, j}(u)  , 
\\
 {\bf  \nu}_{a, \text{ns}}(u) &=& \sum_{j=0}^{\infty} h^{2 j - L } \, {\bf  \nu}_{a, \text{ns}, j}(u),
\eeqa
where ${ \bf p }_i$ is defined by $\bP_i = (h x )^{-L} \, { \bf p }_i$. 
 Naturally, the normal scaling expansion of the $\bf p$'s is simply related to (\ref{eq:ansatz1})-(\ref{eq:ansatz5}) by expanding at weak coupling both the coefficients and the Zhukovsky variable  $x(u) \sim \frac{u}{h} ( 1  + \text{O}(h^2) )$. 
The procedure also computes the normal scaling expansions of $\widetilde{ \bf p }_i$:
\beqa
\widetilde{\bf  p}_{i, \text{ns}}(u) &=& \sum_{j=0}^{\infty} h^{2 j} \, 
\widetilde{\bf  p}_{i, \text{ns}, j}(u). 
\eeqa
 It is important to describe how the analytic properties are modified in the perturbative limit. The branch cuts at $u \in (-2 h, 2 h ) + i \mathbb{Z}$  shrink to zero, and, at every order in the expansion, poles normally appear, correspondingly, at positions  $u \in i \mathbb{Z}$.  
 However, the underlying analytic structure of the QSC imposes some very nontrivial constraints. In particular, we shall use the fact that at finite coupling the combinations
\beqa
&& \nu_{a}(u) +  {\widetilde \nu}_a(u) =  \nu_{a}(u) + \sig \, \nu_a^{[2]}(u) , \label{eq:sing1}\\
 && \frac{\nu_{a}(u) -  {\widetilde \nu}_a(u) }{ \sqrt{ u^2 - 4 h^2 } } = \frac{\nu_{a}(u) -  \sig \,  \nu_a^{[2]}(u) }{ \sqrt{ u^2 - 4 h^2 } }, \label{eq:sing2}
\eeqa 
are free of cuts and analytic on the whole real axis. This means that the normal scaling re-expansion of these combinations should not develop a pole around $u= 0$ at any perturbative order. This set of constraints is used in an essential way in the procedure.
 Besides, with the disappearance of the cuts, there is no direct relation between the values of ${\bf  p}_{i, \text{ns}, k}(u)$ and $\widetilde{\bf  p}_{i, \text{ns}, k}(u)$ at any given order. 
 Nonetheless, the two expansions are nontrivially related through the finite-coupling ansatz (\ref{eq:ansatz1})-(\ref{eq:ansatz5}), since, after the replacement $x \rightarrow { \widetilde x }= 1/x$, the latter yields an expansion for $\widetilde{ \bf P }_i$, which is expected to converge at least in an open neighbourhood of $ |x| = 1$ \cite{Marboe:2014gma}. As we shall see, this will provide precious information for the  perturbative method.  

\subsection{Summary of a single iteration}

Each iteration of the algorithm starts from the knowledge of ${\bf p }_{1, \text{ns}, n}$ and 
 ${\bf p }_{2, \text{ns}, n}$, complemented with limited information on the normal scaling expansion of $\widetilde{ \bf p }_1$ and $\widetilde{ \bf p }_2$,  and allows to compute every other quantity at the same perturbative level. 
 The input needed on $\widetilde{ \bf p }_{1, \text{ns}, n}$ and $\widetilde{ \bf p }_{2, \text{ns}, n}$  consists only in the  coefficients of their Laurent expansions around $u=0$, up to the $\text{O}(1)$   or $\text{O}(u)$ term, respectively.  
 For $n=0$, all this information can be easily extracted from the the ansatz (\ref{eq:ansatz1})-(\ref{eq:ansatz2}), and we find 
 \beqa\label{eq:smallu0}
 { \bf p }_{1,\text{ns}, 0} = 1 , &\;\;\;\;\;&  { \bf p }_{2,\text{ns}, 0} = 0 , \label{eq:LOstart1}\\
 \widetilde{ \bf p}_{1, \text{ns}, 0}(u) \sim 1 + \text{O}(u), &\;\;\;\;\;& \widetilde{ \bf p}_{2, \text{ns}, 0}(u) \sim u + \text{O}(u^2). \label{eq:LOstart2}
 \eeqa
For the reader's convenience, we list below the main steps of the algorithm:

\begin{description}[leftmargin=*]

\item[Step 1:] \emph{  Fix the singular parts of ${\bf p}_0$  and ${\bf p}_3$ }

At a generic iterative order, the first step is to determine the singular part of 
${\bf p }_{0, \text{ns}, n}$ and ${ \bf p }_{3, \text{ns}, n}$. 
Since the $\bP$'s have a single cut on the principal Riemann sheet, in the normal scaling expansion they can develop pole singularities only at $u=0$.
  From the weak coupling expansion of the ansatz (\ref{eq:ansatz3}), (\ref{eq:ansatz5}), we see that, once the pole part is fixed, the regular part around $u=0$ depends only on the coefficients of the polynomials (\ref{eq:polyns}). 
  This is very important in order to keep the  number of unknowns in the algorithm under complete control 
  at every iteration.   

To fix the pole parts, we consider the sum of equations (\ref{eq:nut}) and (\ref{eq:nutt}) for $a=1,2$, taking 
into account (\ref{eq:periodic}). Then we obtain
\beqa
\left(\nu_1 + \sig  \, \nu_1^{[2]} \right) \left( {\bf p}_0 - ( h x )^L \right) &=& { \bf p }_2 \, \left( \nu_2 + \sig \,\nu_2^{[2]} \right) - {\bf p}_1 \, \left( \nu_3 + \sig \, \nu_3^{[2]} \right)   , \label{eq:pole1}\\
\left( \nu_2 + \sig \, \nu_2^{[2]} \right)\left( {\bf p}_0 + ( h x )^L \right)   &=& {\bf p}_3\,  \left(\nu_1 +  \sig \, \nu_1^{[2]} \right) + {\bf p}_1 \, \left( \nu_4 + \sig \, \nu_4^{[2]} \right).
\label{sing2}
\eeqa
 Notice that the combinations in brackets are always of the form (\ref{eq:sing1}), and therefore free of poles at $u=0$ at any order. The order-$\text{O}( h^0)$ values of the ${\bf p}$'s and $h x$ are  also regular at $u=0$. Therefore, the   pole part of ${ \bf p }_{0, \text{ns}, n}(u)$ and ${ \bf p }_{3, \text{ns}, n}(u)$ can  be determined without requiring the knowledge of  $\nu_{a, \text{ns}, n}$. \\
 At leading order, the above computation simply fixes the singular part to be zero. 
\item[Step 2:] \emph{   Solve the Baxter equation for $\nu_1$, with unfixed polynomial part of ${\bf p}_0$ ; then fix the polynomial part using analyticity} 

 Then we consider the first of the Baxter  equations (\ref{eq:TQ}), reading explicitly 
\beqa\label{eq:inhNu1}
\frac{\nu_1^{[3]} }{ \bP_1^{[1]} } - \frac{\nu_1^{[-1]} }{ \bP_1^{[-1]}} - \sig \,\(  \frac{\bP_0^{[1]}}{\bP_1^{[1]}} - \frac{\bP_0^{[-1]}}{\bP_1^{[-1]}}\) \, \nu_1^{[1]} = - \sig \, \(\frac{\bP_2^{[1]}}{\bP_1^{[1]}}-\frac{\bP_2^{[-1]}}{\bP_1^{[-1]}}\)\nu_2^{[1]} .
\eeqa
At leading order in the weak coupling expansion, the rhs can be dropped since $\bP_2 \sim \text{O}( h^2)$, and (\ref{eq:inhNu1})  reduces to the well-known 2-loop Baxter equation :
\beq\label{eq:baxter0}
 T_0(u) Q(u) = 
Q(u+i) \, (u+i/2)^L - Q(u-i) \, (u-i/2)^L, 
\eeq
where $\nu_{1, \text{ns}, 0}(u)\propto Q^{[-1]}(u)$ and 
\beqa
T_0(u) = \sig \,\left( M_{L}^{(0)}(u+i/2) -  M_{L}^{(0)}(u-i/2) \right),
\eeqa
where $M_L^{(0)}(u)$ is the coefficients of $h^{0}$ in the weak coupling expansion of $M_L(u)$.  
It is well-known that the (\ref{eq:baxter0}) admits a discrete set of solutions determining simultaneously the transfer matrix eigenvalue $T_0(u)$ and the Baxter polynomial $Q(u) = \prod_{j=1}^S (u - u_j)$. Physical solutions are also required to satisfy the zero-momentum condition (ZMC) \cite{Minahan:2008hf}, which can be written as
\beqa\label{eq:zmc}
\left(\prod_{j=1}^{S}\frac{u_j+\frac{i}{2}}{u_j-\frac{i}{2}}\right)^2 = 1 \;\;\;\; \leftrightarrow \;\;\;\;
\frac{Q\left(+\frac{i}{2}\right)}{Q\left(-\frac{i}{2}\right)} = \pm 1 . 
\eeqa
The sign ambiguity in (\ref{eq:zmc}) is an important difference as compared to the $\mathcal{N}$=4 SYM case. 
 Notice that the sign  on the rhs of (\ref{eq:zmc}) must be identified with the value of $\sigma$, since  requiring the absence of poles in (\ref{eq:sing2}) at the leading order we obtain precisely $Q^{[1]}(0)/Q^{[-1]}(0) = \sigma$. In summary, picking a Baxter polynomial corresponds to choosing a state in the theory and unambiguously defines the  value of $\sigma$ to be used in all subsequent iterations.  
 We will discuss shortly how to fix the precise proportionality coefficient $\alpha$ in $\nu_{1, \text{ns}, 0}(u) = \alpha \, Q^{[-1]}(u)$. 
 
At higher orders, plugging the previously computed values of ${\bf p}_{1,\text{ns}, k}$, ${\bf p}_{2,\text{ns}, k}$, $k \leq n$, $\nu_{2,\text{ns}, l}$, $l \leq n-1$, together with the pole part of  ${\bf p}_{0 , \text{ns}, n}$ into (\ref{eq:inhNu1}), we are  required to solve an inhomogeneous Baxter equation of increasing complexity. 
 The technique for its solution,  coming from \cite{Marboe:2014gma}, is explained in Appendix \ref{app:inhomo}. 
 The output of the procedure returns $\nu_1$ as a combination of rational functions of $u$ and generalized Hurwitz zeta functions. The solution depends on the still undetermined coefficients of the polynomial $M_{L, n}(u)$, plus an additional set of integration constants $\phi^{\text{per}}_{a, k}$, $\phi^{\text{anti}}_{a, k}$ (see Appendix \ref{sec:perPhi} for details). 

To fix all these parameters, we use the analyticity requirements coming from the regularity of (\ref{eq:sing1}),(\ref{eq:sing2}) for $a=1$, which can be seen as the higher loops analogue of imposing the ZMC and polynomiality for $Q$ and $T_0$ in (\ref{eq:baxter0}). This yields a linear system of equations, which determines all coefficients, apart from  $m_0^{(2n)}$ (which is unfixed due to the symmetries of the system, see Appendix \ref{app:symmetry}) and another coefficient, which is the analogue of $\alpha$ at leading order, and corresponds to the freedom to re-define the solution as $\nu_{1, \text{ns}, n}(u) \rightarrow  \nu_{1, \text{ns}, n} + \alpha_n  \, Q(u)$. 
 We will show in the next step how to resolve this ambiguity. 
 
Notice that the size and complexity of this linear system increases with the order of iteration. This is one of the delicate parts of the algorithm, since the solvability of the linear system requires to use nontrivial relations between the Euler-Zagier sums that are the numbers generated by the procedure and make up the matrix of coefficients. Luckily, a full set of  relations -- reducing any multiple sum to a combination of lower weight zetas from a minimal basis -- have been tabulated up to transcendentality twelve in the datamine of \cite{DataMine}\footnote{In the case of non-alternating sums, these relations are provided up to transcendentality twenty-two. }. For the 12-loop computation, we used these results up to weight nine 
(which also critically improves the program's speed and memory usage); at least one more iteration should be relatively simple to perform making full use of these decomposition rules.

\item[Step 3:] \emph{ Compute $\nu_3$, $\widetilde{ \bf p }_2$ and  ${\bf p }_4$; fix the constants $\alpha_n$ } 

We can now easily  determine $\nu_3$ from the first equation in (\ref{eq:nut}):
\beqa
\sigma \,  \nu_1^{[2]} = \bP_0 \, \nu_1  - \bP_2 \,  \nu_2 + \bP_1 \, \nu_3 ,
\eeqa
 where we know every other term at the relevant perturbative order.  The second equation from the system (\ref{eq:Ptil}):
\beq
{ \widetilde {\bP} }_{2} - { \bf P }_{2} = \sig \, \( \nu_3 {  \nu }_1^{[2]} - \nu_1 { \nu }_3^{[2]} \),
\eeq
can then be used to determine $\widetilde{ \bf p }_2$. 
 Considering the small-$u$ behaviour of $\widetilde{ \bf  p }_{2, \text{ns}, n }(u)$, and matching it with the small-$u$ expansion predicted from the previous iteration (see Step 6), it is possible to fix the coefficient $\alpha_n$. In particular, at leading order, the constraint (\ref{eq:LOstart2}) allows us to fix\footnote{ The sign of the square root in (\ref{eq:alpha0})  is unimportant as the QSC is symmetric under $\nu_a \rightarrow -\nu_a$.}
 \beqa\label{eq:alpha0}
 \alpha = \frac{h^{ -L } }{ \left( 2 \, (Q^{[1]})^2 \, \left. \partial_u \log \frac{ Q^{[1]}}{Q^{[-1]} } \right|_{u=0} \right)^{\frac{1}{2}}}.
 \eeqa 
Finally, we compute ${\bf p }_4$ using the quadratic constraint (\ref{eq:quad}). Notice that, since ${ \bf p}_{2, \text{ns}, 0} \sim \text{O}(h^2)$, this only requires to know ${ \bf p}_3$ at the previous perturbative order.

\item[Step 4:]
 \emph{  Solve the Baxter equation for $\nu_2$, with unfixed coefficients for the polynomial part of ${\bf p }_3$; impose the analyticity constraints }
 
We then consider the second of the Baxter equations (\ref{eq:TQ}):
\beqa\label{eq:inh2}
\frac{\nu_2^{[3]} }{ \bP_1^{[1]} } - \frac{\nu_2^{[-1]} }{ \bP_1^{[-1]}} +  \sig \,\(  \frac{\bP_0^{[1]}}{\bP_1^{[1]}} - \frac{\bP_0^{[-1]}}{\bP_1^{[-1]}}\) \, \nu_2^{[1]} =  \sig \, \(\frac{\bP_3^{[1]}}{\bP_1^{[1]}}-\frac{\bP_3^{[-1]}}{\bP_1^{[-1]}}\)\nu_1^{[1]} .
\eeqa
 Notice that, even at the first iteration, the equation is inhomogeneous. Besides, its homogeneous part differs from the Baxter equation (\ref{eq:baxter0}) for a sign in front of the transfer matrix eigenvalue.\\ 
 As explained in Appendix \ref{app:inh2}, we can solve this equation and determine  $\nu_{2, \text{ns}, n}$ as a function of the yet unfixed coefficients of the polynomial part of ${ \bf p }_3$. 
 By requiring the regularity of the combinations (\ref{eq:sing1}) and (\ref{eq:sing2}) for $\nu_{2, \text{ns}, n}$, one again finds a system of linear equations whose solution fixes all of the coefficients of the polynomial $K_{2 L + 1}$ at order $h^{2 n}$, apart from the three coefficients $k_0^{(2n)}$, $k_L^{(2n)}$, $k_{2 L }^{(2n)}$, which are left free due to the symmetries of the system (see Appendix \ref{app:symmetry}), and the coefficient of the highest power $A_3^{(2n)}$. The latter contains the anomalous dimension, and will be determined in the next step.

\item[Step 5:] \emph{ Compute  $\widetilde{ \bf p }_{1, \text{ns}, n}$  and fix the anomalous dimension; compute $\nu_4$ }

In terms of the previously found solution for $\nu_2$, we  can compute $\widetilde{ \bf p}_{1, \text{ns}, n}$ using the first equation of the system (\ref{eq:Ptil}):
\beq\label{eq:P1t}
{ \widetilde {\bP} }_{1} - { \bf P }_{1} = \sig \, \(\nu_2 {  \nu }_1^{[2]} - \nu_1 { \nu }_2^{[2]} \).
\eeq
 
 By matching the expected leading behaviour at $u \sim 0$, we finally fix the last remaining coefficient $A_3^{(2n)}$, and determine the relevant term of the anomalous dimension $\gamma_{2n+2}$ from (\ref{eq:A2A3}).
 
 At this stage we can also compute $\nu_{4, \text{ns}, n}$ by inverting the second
equation of (\ref{eq:nut}):
\beqa
\sigma \, \nu_2^{[2]} =  -\bP_0 \, \nu_2 + \bP_3\,  \nu_1  + \bP_1 \, \nu_4 .
\eeqa
 
\item[Step 6:] \emph{ Compute the seed for the next iteration }

Finally, starting from the knowledge of $\left\{ \widetilde{\bf p}_{a, \text{ns}, n} \right\}_{j=0}^n$, we shall show how to determine the quantities needed for the next iteration, namely ${\bf p}_{a,\text{ns},n+1}$ and the small-$u$ behaviour of $\widetilde{\bf p}_{a, \text{ns}, n+1}$ up to the same powers as in (\ref{eq:LOstart2}) ($a=$1,2). 
 
 To achieve this, we start by noticing that the ansatz (\ref{eq:ansatz1})-(\ref{eq:ansatz5}), which underlines the whole method, is naturally organised in terms of the parameter $y = h/x$. 
 Besides the normal scaling expansion, it is very convenient to consider also the \emph{double scaling} expansion, obtained by expanding 
 ${\bf p}_1$ and ${ \bf p}_2$ at fixed value of $y$. We can summarise the difference between the two expansions by writing
\begin{align}
&\text{normal scaling: }& \;\;\; &{\bf p }_{a} = { \bf p}_a\left( \frac{h}{x(u ; h )} , h \right)  , \; &h \sim 0 , \;\;\; u\text{ fixed }, \\
&\text{double scaling: }& \;\;\; &{\bf p }_{a} = { \bf p}_a(y , h )  , \; &h \sim 0 , \;\;\; y\text{ fixed }.
\end{align}
 The seed quantities for the $n+1$-th iteration will be computed from the truncated double scaling series:
 \beqa\label{eq:trunc}
 \mathcal{S}_{a,n}(y) = \sum_{j=0}^{ n } h^{2 j } \,  { \bf p }_{a, \text{ds} , j }(y) , \; a=1,2.
 \eeqa
Let us then show how to obtain (\ref{eq:trunc}) order by order. 
  The main observation is that, setting $y \rightarrow h/\widetilde{x}(u) = h x(u)$ in (\ref{eq:trunc}) and 
  re-expanding  at normal scaling $ h x(u) \sim u + \text{O}(h^2)$, one should find precisely the truncated  series for $\widetilde{ \bf p}_{a, \text{ns}}(u)$, up to the same order. 
 The precise relation between the two expansions is\footnote{Since $\mathcal{S}_{-1}=0$, at the first iteration  (\ref{eq:dsEq}) is simply
\beqa\label{eq:dsEq0}
 { \bf p }_{a, \text{ds} , 0}(y ) = \left. \widetilde{ \bf p }_{a, \text{ns},0 }(u) \right|_{ u \rightarrow  y} , \;\;\; a = 1, 2 . \nn
 \eeqa
} \cite{Marboe:2014gma}:
 \beqa\label{eq:dsEq}
 { \bf p }_{a, \text{ds} , n }(y ) = \left. \widetilde{ \bf p }_{a, \text{ns}, n }(u) \right|_{ u \rightarrow  y} - \left. { \bf p }'_{a, \text{ns}, n}( u ) \right|_{ u \rightarrow  y} , \;\;\; a = 1, 2 ,
 \eeqa
where ${ \bf p }'_{a, \text{ns}, n }(u)$ is the coefficient of $h^{2n}$ in the normal scaling  expansion of $\mathcal{S}_{n-1}( h x(u) )$. 

Relation  (\ref{eq:dsEq}) is used in the program to compute $\mathcal{S}_{a,n}(y) = \mathcal{S}_{a, n-1}(y) + { \bf p }_{a, \text{ds} , n }(y )$. Now, since $h/x(u) \sim \text{O}(h^2)$, it is enough to re-expand  $\mathcal{S}_{a, n}(  h/x(u) )$  at fixed $u$ to obtain ${\bf p}_{a, \text{ns}, n+1}(u)$. 

 Finally, 
notice that the ansatz (\ref{eq:ansatz1})-(\ref{eq:ansatz5}) implies that  the small-$y$ behaviour of the double scaling expansion is, for every $l > 0$, 
\beqa
&&{ \bf p }_{1, \text{ds} , l}(y) \sim \text{O}(y), \;\;\;\; { \bf p }_{2, \text{ds} , l}(y) \sim  \text{O}(y^2) , 
\eeqa
and from (\ref{eq:dsEq}) we see that
\beqa\label{eq:dsEq3}
&&\widetilde{ \bf p }_{1, \text{ns}, n+1 }(u) -  { \bf p }'_{1, \text{ns}, n+1}( u )  = \text{O}(u), \\
&&\widetilde{ \bf p }_{2, \text{ns}, n+1 }(u) -  { \bf p }'_{2, \text{ns}, n+1}( u )  = \text{O}(u^2),
 \eeqa
where ${ \bf p }'_{a, \text{ns}, n+1}( u )$ can be computed from the expansion of $\mathcal{S}_{a,n}( h x(u) )$. This relation enables us to fix the small-u behaviour of $\widetilde{ \bf p }_{1, \text{ns}, n + 1 }(u)$, $\widetilde{ \bf p }_{2, \text{ns}, n + 1 }(u)$, up to and including the $\text{O}(u^0)$, $\text{O}(u)$ terms, respectively, as needed to restart the algorithm. 
 
\end{description}

\section{Testing the results}\label{sec:results}

The simplest unprotected operator belonging to the symmetric sector of ABJM is the {\bf 20} with $su(2)$ and $sl(2)$ representatives with charges $L=2, S=1$ and $L=1, S=1$ respectively. 
By iterating the procedure illustrated in Section \ref{sec:algo}, we obtained the  12-loop result reported in Appendix \ref{app:results}. In the same Appendix and in an attached notebook we also list the dimensions of a few other operators. 

Up to 4 loops, we can directy compare these results with the literature  \cite{Beccaria:2009ny,Minahan:2009aq,Minahan:2009wg,Leoni:2010tb,Beccaria:2010kd}. In particular, we tested our method for $L=1$ and several values of $S$ ($S=1,...,20$) 
against the formula of \cite{Beccaria:2009ny,Beccaria:2010kd}, 
\beqa
\gamma^{L=1,S} &=&4 \,h^2(H_1+H_{-1})-16\, h^4 {\Big (}H_{-3} -H_3 +H_{-2,-1} -H_{-2,1} +H_{-1,-2} -H_{-1,2} -H_{1,-2}  \nonumber\\
&+&H_{1,2} -H_{2,-1} +H_{2,1} +H_{-1,-1,-1} 
-H_{-1,-1,1} -H_{1,-1,-1} +H_{1,-1,1}  {\Big )} \nn\\
&+&  4 \, h^4 (H_1 -H_{-1} )\,\mathcal{W}(1, S) + \text{O}(h^6) ,
\eeqa 
where we have used the shorthand notation  $H_{a_1, \dots, a_k} \equiv H_{a_1, \dots, a_k}(S)$ for the generalized harmonic numbers, and the wrapping term $\mathcal{W}(L, S)$ was calculated in 
\cite{Beccaria:2009ny}:
\beqa
\mathcal{W}(L, S)&=&-\frac{i}{2} \,\sum_{ M=1}^{\infty}(-1)^M\mathop{Res}_{q=i M}\frac{4^L Q\left(\frac{q-i(M-1)}{2}\right)}
{(q^2+ M^2)^L Q\left(\frac{q-i(M+1)}{2}\right)Q\left(\frac{q+i(M-1)}{2}\right)Q\left(\frac{q+i(M+1)}{2}\right)}\nonumber\\
&\times&\sum_{j=0}^{M-1}\left[\frac{Q\left(\frac{q-i(M-1)+2ij}{2}\right)}
{Q\left(\frac{q-i(M-1)}{2}\right)}\right]^2 \, \left(\frac{1}{2j-iq-M}-\frac{1}{2(j+1)-iq-M}\right).
\label{WLS}
\eeqa
 The Baxter function to be used in the formula above for $L$=1 states is 
\beq
Q(u)= {}_2 F_1 (-S,iu+1/2;1;2).
\eeq
 The 6-loop contribution, instead, is a new prediction. However, since it is  still not affected by double-wrapping (that sets in starting from eight loops for twist-1 operators\footnote{For a state with twist $L$,  single-wrapping terms start to appear perturbatively at  order $h^{2(L+1)}$, while double-wrapping phenomena are expected to contribute from order $h^{4(L+1)}$. }), we can test it against the NLO expansion of L\"uscher's formula, which is discussed below. 

 In the case of twist-2 operators, we checked our results, up to four loops, against the ABA prediction  \cite{Beccaria:2009ny,Beccaria:2010kd}:
\beqa
\gamma^{L=2,S} &=&4\,h^2(H_1-H_{-1}) + 8 \, h^4{ \Big ( }2H_{-3}+2H_3-H_{-2,-1}-H_{-2,1}-2H_{-1,-2} \nn\\
&&-2H_{-1,2}-2H_{1,-2}-2H_{1,2}-2H_{2,-1}-H_{2,1}{ \Big )}.
\eeqa
At 6 loops we have the first appearance of single-wrapping effects, calculated in \cite{Beccaria:2010kd} for $S=2,4,...,30$ by using the formula (\ref{WLS}) with $L$=2 and
\beq
Q(u)={}_3 F_2(-S/2,iu+1/2,-iu+1/2;1,1;1) ,
\eeq
while the 6-loop ABA result is too long to be written here, but it can be found in \cite{Beccaria:2009ny}.
In order to check the 8-loop result, we need to expand further the L\"uscher terms, as for the twist-1 6-loop term, involving corrections to the rapidities and the first contribution
from the transcendental part of the dressing factor: this will be done in  Section \ref{sec:Luscher}.
We found perfect agreement with all these results.

\subsection{More checks with L\"uscher corrections}
\label{sec:Luscher}
Single-wrapping effects can be calculated  at all orders using L\"uscher's method \cite{Luscher:1985dn}, first generalized in the case of $\mathcal{N}$=4 SYM  in \cite{Janik:2007wt,Bajnok:2008bm}. 
 We shall use the following L\"uscher-like formula, introduced for ABJM by \cite{Bombardelli:2008qd, Lukowski:2008eq}:
\beq
\delta \gamma=-\sum_{M=1}^{\infty}\int\frac{dq}{2\pi}e^{-L\tilde E_M(q)}\sum_{b=1}^4(-1)^{F_b}\prod_{i=1}^S (S^{M,1}_{AA}S^{M,1}_{AB})_{b3}^{b3}(q,p_i),
\label{luscher}
\eeq
where $S_{AA}$ and $S_{AB}$ are the $SU(2|2)$-invariant S-matrices found in \cite{Ahn:2008aa}, $F_b=0, 1$ for bosonic  or fermionic indexes, respectively, and $\tilde E_M(q)$ is the mirror energy
\cite{Wrapping,Arutyunov:2007tc}
\beq
\tilde E_M(q)=2 \, \mathrm{arcsinh} \left(\frac{\sqrt{q^2+M^2}}{4h}\right).
\eeq
Using the techniques developed in \cite{Bajnok:2008bm} for the ${\cal N}$=4 SYM case, it is possible to calculate the integrand appearing in 
(\ref{luscher}) at various orders in the weak coupling expansion.
In particular, in the symmetric case $A=B$ in (\ref{luscher}), we can use S-matrix elements squared with respect to those used in \cite{Bajnok:2008bm} and the same (fused) 
scalar factor $S_{sl(2)}^{M,1}(q,p_i)$.
 At the leading order at weak coupling, (\ref{luscher}) reduces to (\ref{WLS}) for given $L$ and $S$.  More interestingly, expanding (\ref{luscher}) to higher powers of $h$, we performed further nontrivial checks of the QSC-based computation.  
  As discussed in \cite{Bajnok:2008bm}, to go beyond the leading order it is essential to consider also the finite-size corrections to the physical rapidities or momenta, which yield an additional contribution to the anomalous dimension 
\beq
\delta \gamma_{ABA}=\sum_{k=1}^S \gamma_{ABA}'(p_k)\delta p_k .
\label{Emod1}
\eeq
 The asymptotic momenta $p_i$ appearing in (\ref{luscher}) and (\ref{Emod1}) are fixed by the Asymptotic Bethe Ansatz equations
\beq
\left(\frac{x_k^+}{x_k^-}\right)^{L}=-\prod_{\stackrel{j\neq k}{j=1}}^S\frac{x_k^- - x_j^+}{x_k^+ - x_j^-}\frac{1-1/{x_k^+ x_j^-}}{1-1/{x_k^- x_j^+}} \, e^{2i\Theta(x^{\pm}_k,x^{\pm}_j)},
\label{ABA}
\eeq
where
 $\Theta(x^{\pm}_k,x^{\pm}_j)$ is the BES dressing phase \cite{Beisert:2006ez}, $x_i^{\pm}=x(u_i \pm i/2 )$ and $p_i = -i \log \frac{x_i^+}{x_i^-}$, while the corrections $\delta p_k$'s are determined by the system \cite{Bajnok:2008bm} 
\beq
L\delta p_k +i\sum_{j=1}^S\delta p_j\partial_{p_j}\log S(p_k,p_j) = \delta\Phi_k ,
\eeq
where $S(p_k,p_j)$ is the scattering phase involved in the corresponding ABA equation, and
\beq
\delta\Phi_k=\sum_{M=1}^{\infty}\int\frac{dq}{2\pi}e^{-L\tilde E_M(q)}\sum_{b=1}^4(-1)^{F_b}\prod_{\stackrel{j=1}{j\neq k}}^S (S^{M,1}_{AA}S^{M,1}_{AB})_{b3}^{b3}(q,p_j)
\partial_q(S^{M,1}_{AA}S^{M,1}_{AB})_{b3}^{b3}(q,p_k) .
\label{Emod2}
\eeq

 In the case $L=1$, $S=1$,  the corrections in (\ref{Emod1}) vanish, since 
the rapidity is fixed at $u=0$ by the trivial Bethe equation $( {x^{[1]}}/{x^{[-1]}})^L=-1$, and 
 the contribution to the energy determined by the ABA is exactly $\gamma^{L=1,S=1}_{ABA}=\sqrt{1+16h^2}-1$ at any order. 
 The only complication at 6 loops comes from the transcendental part of the dressing factor. The latter however can be treated analytically  using the techniques explained in \cite{Bajnok:2009vm}. Finally, one gets
\beq
\gamma_{6,wrapping}^{L=1,S=1}=288 \, \zeta_3-768+ 320 \, \zeta_2 + \frac{504  }{5} \, \zeta_2^2  + 384 \, \zeta_2\,\zeta_{-1},
\eeq
while $ \gamma_{6,ABA}^{L=1,S=1}=256$, matching exactly the result reported in Appendix \ref{app:results}.  \\
We also checked the 6-loop term in the $L=1, S=2$ case, for which we had to calculate the corrections (\ref{Emod1})-(\ref{Emod2}) to the asymptotic momenta
\beq
p_1=-p_2=\frac{\pi}{2}-2h^2+12h^4+\text{O}(h^6).
\label{momentumL1}
\eeq
In particular, we evaluated the integrals appearing in (\ref{luscher}) and (\ref{Emod2}) using \texttt{NIntegrate} in \emph{Mathematica},  with \texttt{WorkingPrecision} (\texttt{WP}) set to 50, and summed them up to $M=2000$, obtaining
\beq
\gamma_{6,wrapping}^{L=1,S=2}=47.234481782125339488\dots . 
\label{L1S26loop}
\eeq
Once the asymptotic energy 
$\gamma_{6,ABA}^{L=1,S=2}=224$ is added, this agrees with the analytic result reported in Appendix \ref{app:results} up to 18 digits.  To improve the accuracy by reducing the errors coming from the truncation of the sums, we employed  the technique described in Section 5 of \cite{Bajnok:2012bz}. First, we derived a large-$M$  approximation for the (symmetrised) integrands appearing in (\ref{luscher}) and (\ref{Emod2}). After rescaling $q\rightarrow sM$ and keeping the first six orders, one finds 
\beqa
&&\hspace{-1cm}\mbox{integrand}_{L,S=2}(s,M)\sim (-1)^M \sum_{n=0}^5\frac{1}{M^{4 + 2 L +2n}}\left(a_n^{L,S=2}(s)+b_n^{L,S=2}(s)\log\frac{1}{M}\right) . 
\eeqa
 The contribution of the $M>2000$ terms can then be estimated by integrating the functions $a_n(s)$ and $b_n(s)$ and performing the sum from $M=2001$ to $\infty$. Adding this correction, the agreement with the analytic result coming from the QSC method reaches 48 digits. 
 Such a high precision is sufficient, using EZ-Face \cite{EZ-face}, to recover the rational coefficients of the zeta functions appearing in the QSC result.

Starting at order $h^8$, L\"uscher's formula (\ref{luscher}) fails to give the complete answer for twist-1 operators, due to the  appearance of  double-wrapping
effects. However, in ${\cal N}$=4 SYM one has that  the  highest transcendentality part of the result is affected only by single wrapping at any coupling  \cite{Bajnok:2012bz, leurent2013multiple}. Based on this expectation, we computed the ABA and single-wrapping contribution to the maximally transcendental part of the anomalous dimension, at 8 loops and beyond. In the present case, this does not match the result of the perturbative solution of the QSC, showing, as anticipated in the Introduction, that these terms also include double wrapping corrections. To gain more insights, it would be interesting to extend the L\"uscher-like techniques to the double-wrapping order, using the method developed in \cite{Bombardelli:2013yka}.

For twist-2 operators, instead, double wrapping effects  start only at 12 loops, and we can still test the 8- and 10-loop results. In particular, for $S$=2 we obtained
\beq
p_1=-p_2=\frac{\pi}{3}-\frac{h^2}{\sqrt{3}}+\frac{23 h^4}{6\sqrt{3}}-\frac{8h^6(20+9 \, \zeta_3)}{9\sqrt{3}}+\frac{h^8(9857+5760\,\zeta_3+8640\,\zeta_5)}{108\sqrt{3}}+\text{O}(h^{10}),
\label{momentum}
\eeq
from the perturbative expansion of the Bethe equations (\ref{ABA}).
Then the ABA contribution to the energy is
\beqa
\label{EABA}
 \gamma_{ABA}&=&\sum_{k=1}^2\sqrt{1+16  h^2\sin(p_k)^2}-1\\
&=&4h^2-8h^4+32h^6-4(39+8 \, \zeta_3)h^8+4(209+72 \, \zeta_3+80 \, \zeta_5)h^{10}+\text{O}(h^{12}) . \nn
\eeqa
Plugging the solutions (\ref{momentum}) into (\ref{luscher}) and expanding   at leading order, we get the expected result
\beq
\gamma_{6,wrapping}^{L=2,S=2}=-\frac{16}{3}+\frac{64}{3} \, \zeta_2-42 \, \zeta_4 .
\eeq
At 8-loop, evaluating numerically with \texttt{WP}=50 and summing up to $M=2000$ the integrals coming from the expansion of (\ref{luscher}) and (\ref{Emod2}), we obtain
\beq
\gamma_{8,wrapping}^{L=2,S=2}=302.916652975526460098546231 \dots , 
\label{L2S28loop}
\eeq
that, added to the corresponding terms of (\ref{EABA}), agrees with the result reported in Appendix \ref{app:results} up to 25 digits. A 49 digits agreement is obtained by adopting the asymptotic resummation method described above. 

Since the complexity and the size of the integrands involved in these calculations increase drastically with the value of $S$ and the loop order, we managed to perform
just some lower-accuracy checks for the $L=2$, $S=2$ operator at 10 loops and the $L=1$, $S=3,4$ operators at 6 loops. Keeping the first 200 terms in the sums, we found 
\begin{align}
\gamma_{10,wrapping}^{L=2,S=2}=-4711.383655826655401\dots, \gamma_{6,wrapping}^{L=1,S=3}=-35.71109708632629843\dots , 
\end{align}
which both agree with the  results reported in Appendix A up to 17 digits, once we add $\gamma_{10,ABA}^{L=2,S=2}=4 (209 + 72 \zeta_3 + 80 \zeta_5)$ and 
$\gamma_{6,ABA}^{L=1,S=3}=82720/243$ respectively. Finally, including the first 50 integrals for the case $L=1$, $S=4$, we obtained 
$\gamma_{6}^{L=1,S=4}=343.369048487641320\dots $, matching the expected result with 16 digits precision.

 \subsection{Comparison with TBA}
 \begin{figure}[h!]
 \centering
 \includegraphics[width=10cm]{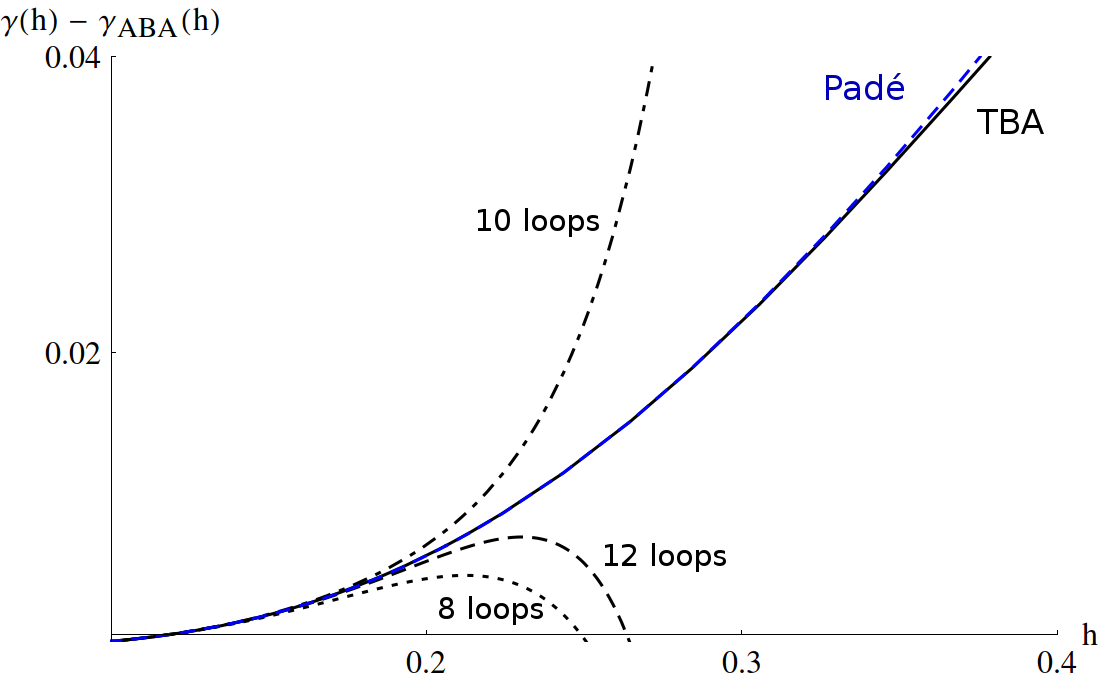}
 \caption{ Comparison of perturbative results with the TBA data of \cite{LevkovichMaslyuk:2011ty}. 
 Notice the excellent agreement of the non-perturbative TBA result with the $\left[6/6\right]$ Pad\'e approximant curve obtained from the 12-loop dimension. \label{fig:TBA}
 }
\end{figure}
 The exact anomalous dimension of the ${ \bf 20 }$ operator at finite values of $h$ was studied by F. Levkovich-Maslyuk in \cite{LevkovichMaslyuk:2011ty}, up to values of $h \sim 1$, by solving numerically the TBA equations proposed in \cite{Gromov:2009at}. In  Figure \ref{fig:TBA} is shown the interpolation of the TBA data of \cite{LevkovichMaslyuk:2011ty}, together with various truncations of the perturbative result,
\beqa
\gamma^{L=1,S=1}(h) = 8 \, h^{ 2} -26.3189 \, h^{ 4} +195.487 \, h^{  6} -1966.44 \, h^{ 8} + 22309.5 \, h^{ 10} -271422. \, h^{12} + \dots \nn
\eeqa
The plots show the difference between the anomalous dimension and the exact ABA result $\gamma^{L=1,S=1}_{\text{ABA}} = \sqrt{1 + 16 \, h^2 }-1$. 
 An order  $\left[6/6\right]$  Pad\'e extrapolation of the 12-loop result yields
 \beqa
\text{Pad\'e}\left[ \gamma^{L=1,S=1}(h) - \gamma^{L=1,S=1}_{\text{ABA} }(h) \right] =  \frac{38.8151 \,h^{ 6} +5.6811 \,h^4 }{164.024 \,h^6+81.7529 \,h^4+17.4840 \, h^2 + 1.},
 \eeqa
which is in remarkably good agreement with the TBA data up to $h \sim 0.4$, as can be seen in the figure. This goes  beyond the radius of convergence of the perturbative series, which -- by analyticity arguments  supported by a fair amount of numerical evidence in $\mathcal{N}$=4 SYM \cite{Beisert:2006ez,Volin:2008kd, Marboe:2014gma} -- is expected to be $|h| < \frac{1}{4}$. 
The approximation is still rather good decreasing the number of loops, e.g., with the optimal choice of the order of the Pad\'e approximant, the relative error in the estimate of $| \gamma(h) - \gamma_{\text{ABA} }(h)|$, compared to the TBA value at $h \sim 0.3$, is $2.4 \times 10^{-2}$, $1.2 \times 10^{-2}$, $2.1 \times 10^{-3}$ for the Pad\'e extrapolation of the 8-, 10-, and 12-loop results, respectively. 
 
\section{Conclusions}\label{sec:conclusions}
The discovery of integrability in the context of the AdS/CFT correspondence has  opened the exciting opportunity to solve certain interacting gauge theories in 3d and 4d. Within the powerful integrability setup, exact expressions for particular  physical  observables are often fairly easy to obtain, revealing the beautiful simplicity of the final outcomes. One of the recent achievements  in this research  field is related to  the study of  anomalous dimensions  
and their weak coupling  expansion  through  the so-called Quantum Spectral Curve. 

In this article, relying  on the success obtained  for ${\cal N}$=4 SYM,   we have developed an algorithm to solve perturbatively 
the  QSC for the  ABJM model, finding a perfect agreement both with the TBA numerical results and with earlier 
perturbative computations and predicting several new terms. We are planning to optimise the \emph{Mathematica} code and to render it publicly available. 

Along the way, we have recorded  some important formal differences between the  present case and $\mathcal{N}$=4 SYM. These include
the appearance, for the ABJM model, of  Euler-Zagier sums with both positive and negative signs,  MZVs of even arguments $\zeta_{2n}$ and the experimental observation that the highest transcendentality part of the anomalous dimensions is not completely determined by the ABA and single wrapping corrections but also contains double (and possibly higher) wrapping contributions.
  
There are a few   interesting open  problems related to the current  work. First, as already mentioned in the Introduction, it would be interesting to understand better the structure of the perturbative outcomes, and investigate whether some patterns emerge. 
 One could also try to use our results to deduce expressions for the anomalous dimensions for all values of the spin, as was done in \cite{Marboe:2014sya} in the case of $\mathcal{N}=4$ SYM. This would be particularly interesting for the purpose of exploring various conjectures on the large-spin behaviour \cite{Beccaria:2009ny, Alday:2015eya}. 
   The generalization of this perturbative approach to other sectors of the theory is another obviously important problem. Although  we do not have yet concrete results to discuss,  the generalization to the full 
$OSp(2|2)$ sector should be fairly straightforward, while setting up the iterative procedure for generic operators  appears to be more involved. 
 Finally, it would be very nice to complement the current results by adapting to ABJM the numerical technique  developed  in \cite{Gromov:2015wca} for the non-perturbative solution of the QSC, or to try to transfer some of these  powerful  methods to the study of the QSC for the Hubbard model \cite{Cavaglia:2015nta}.

\medskip
\noindent{\bf Acknowledgments }

We thank Zoltan Bajnok, Francis Brown, Martina Cornagliotto, Davide Fioravanti, Nikolay Gromov, Fedor Levkovich-Maslyuk, Massimo Mattelliano,  Stefano Negro, Simone Piscaglia, Oliver Schnetz and Christoph Sieg for interesting discussions, suggestions and/or past collaboration on related topics. 
We are especially grateful to Christian Marboe and Dmytro Volin for very valuable suggestions and feedback and for sharing with us an early version of their paper \cite{Marboe:2014gma}.

This project was partially supported by the INFN grants FTECP  and GAST and the research grants HOLOGRAV  and 
UniTo-SanPaolo  Nr TO-Call3-2012-0088 {\it ``Modern Applications of String Theory'' (MAST)}.

\vspace{0.3cm}

\appendix

\section{Sample results}\label{app:results}
 In this Appendix we present some explicit results, which can also be found (including three more operators) in the \emph{Mathematica} notebook { \tt Results.nb} attached to the present paper. 

 Using reduction formulas such as the ones in \cite{Broadhurst:2014fga}, the result can be written in terms of a small number of non-reducible sums. We will use the basis of \cite{Broadhurst:1996az,Broadhurst:2014fga}, which is conjectured to be minimal. 
 We find empirically that the $l$-loop result involves only  the basis elements $\zeta_{a_1, \dots, a_k}$ with weight $w = \sum_{i=1}^k  | a_i| \leq l-2$ and depth $k \leq l- 1 - w $. 
 In particular, the 12-loop result can be written in terms of the following sums:
{ 
\beqa
&& \zeta_{-1} , \;\;\;\; \zeta_2 , \;\;\;\; \left\{\zeta_{2 n +1} \right\}_{n=1}^4, \;\;\;\; \dots  \nn \\
&& \zeta_{1, -3} , \;\;\;\; \zeta_{1, -5} ,   \;\;\;\; \zeta_{3,-5},  \;\;\;\;  \zeta_{1,-7} ,\;\;\;\; \dots  \nn\\
&& \; \zeta_{1,1,-3}, \;\;\;\; \zeta_{1,1,-5},  \;\;\;\; \zeta_{1,3, -3}, \;\;\;\; \dots \nn \\
&&\zeta_{1,1,1, -3} , \;\;\;\;  \dots \nn \\
&& \dots \nn
\eeqa
}
We report the results in terms of the interpolating function $h(\lambda)$. Assuming the validity of the conjecture of \cite{Gromov:2014eha} for this quantity, namely,
\beqa\lambda = \frac{\sinh (2 \pi  h(\lambda) ) }{ 2 \pi }\, _3F_2\left(\frac{1}{2},\frac{1}{2},\frac{1}{2};1,\frac{3}{2};-\sinh ^2(2   \pi h(\lambda) )\right),
\eeqa 
the anomalous dimensions can be rewritten in terms of the true coupling constant $\lambda$ using  the expansion:
\begin{align}\label{eq:hlambda}
h(\lambda)^{\text{\cite{Gromov:2014eha}}} 
 &= \lambda \, { \Big ( } 1 -2 \, (\lambda ^2 \, \zeta_2 ) + 15 \, (\lambda^2  \, \zeta_2 )^2 - \frac{5358 }{35}\, (\lambda^2 \, \zeta_2 )^3 +\frac{50501}{28} \, ( \lambda^2 \, \zeta_2)^4  - \frac{88949769}{3850} \, (  \lambda ^{2} \, \zeta_2 )^5   { \Big ) } \nn\\
&+ \dots,
\end{align}
the first  two orders of which have been verified by direct calculations  \cite{Minahan:2009aq}. 
\subsection*{Twist-1} 
 For the {\bf 20} operator with $L$=1, $S$=1,  $Q(u) = u$,  we found:
{ \small
\beqa
&&\gamma_{L=1, S=1} = h^{{\bf {2} }} \, 8 - h^{{ \bf 4 }} \, 16 \, \zeta_2 + h^{{ \bf 6 }} \, \left(\frac{504 \, \zeta_2^2 }{5} + 288 \, \zeta_3  +384 \,\zeta_2 \,\zeta_{-1} + 320 \,  \zeta_2 -512\right) \nn \\
 &+& h^{{ \bf 8 }} \,  { \Big ( } \GamOne^{(8)}_{S=1,2} - \frac{ 12736 \, \zeta_2^2 }{5} -4736 \zeta_3 - 1536 \, \zeta_2 \, \zeta_{-1} -512 \, \zeta_2 + 8192{ \Big ) } \nn\\
   &+& h^{{ \bf 10 }} \,  { \Big ( }  \GamOne^{(10)}_{S=1,2} +\frac{672992 \, \zeta_2^3}{35} + 13632 \, \zeta_5 + 60928 \, \zeta_2 \, \zeta_3 +\frac{287232}{5} \,  \zeta_2^2 \, \zeta_{-1}  +  12288 \, \zeta_{1,-3} + 21504 \,\zeta_3 \, \zeta_{-1} \nn\\
   &+&35840 \, \zeta_2^2 +6144 \, \zeta_2 \,\zeta_{-1}^2 - 61440 \, \zeta_2 \, \zeta_{-1} - 25600 \, \zeta_3- 102400 \, \zeta_2 - 32768 { \Big ) }  \nn\\
&+& h^{{ \bf 12 }} \,  { \Big ( }   \GamOne^{(12)}_{S=1,2} -\frac{559684032 \,\zeta_2^4 }{3325} +\frac{11225408
  \, \zeta_7 }{17}  + \frac{1509696 \, \zeta_2^2 \, \zeta_3 }{85} -\frac{25762912 \, \zeta_2 \, \zeta_5}{17} - \frac{2770944}{7} \, \zeta_2^3 \, \zeta_{-1} \nn\\
&-&245760 \, \zeta_2 \, \zeta_{1,-3}  +  122880 \,\zeta_{1,-5} - 682496 \, \zeta _3^2 - 155136 \, \zeta_5 \, \zeta_{-1}  - 1523712 \, \zeta_2 \,\zeta _3 \,\zeta_{-1}-\frac{39740672 \, \zeta_2^3 }{105} \nn\\
   &-&\frac{1701888}{5}
   \,\zeta_2^2 \,\zeta_{-1}^2  -98304 \, (\zeta_{1,-3} \, \zeta_{-1} -  \zeta_{1,1,-3}) + 145920 \zeta_5 - 1164288  \,\zeta_2 \, \zeta_3  -86016 \,\zeta_3 \, \zeta_{-1}^2\nn\\
   &-& 16384 \, \zeta_2  \, \zeta_{-1}^3 - \frac{73728}{5} \, \zeta_2^2  \, \zeta_{-1} +    688128 \,\zeta_{1,-3} + 1204224 \,\zeta_3 \, \zeta_{-1} +\frac{2400256 \, \zeta_2^2 }{5} \nn\\
   &+&344064 \,\zeta_2 \, \zeta_{-1}^2 +2441216 \, \zeta_3 + 1769472  \, \zeta_2  \,\zeta_{-1} +2490368 \, \zeta_2  -1572864 { \Big )}  +   \text{O}(h^{{ \bf 14 }} ) , \nn
\eeqa
}
 and for $L$=1, $S$=2,  $Q(u) = u^2 -\frac{1}{4}$:
  { \small
\beqa
&&\gamma^{L=1, S=2} =  h^{ \bf 2} \, 8 -  h^{\bf 4} \,  \left(\frac{32}{3} + 16 \,\zeta_2 \right) +  h^{\bf 6} \,\left(    \frac{504 \, \zeta_2^2 }{5}  +  288 \, \zeta_3+ 384 \,\zeta_2 \,\zeta_{-1} +\frac{ 320 \, \zeta_2 }{3}   -\frac{256}{3} \right) \nn\\
   &+& h^{ \bf 8 }{ \Big ( } \GamOne^{(8)}_{S=1,2}-\frac{ 21184 \, \zeta_2^2 }{15} - 1792 \, \zeta_3 + \frac{4352 \, \zeta_2 }{9} +\frac{29696}{27} { \Big ) } \nn\\
   &+& h^{{ \bf 10 }} \,  { \Big ( }   \GamOne^{(10)}_{S=1,2} +\frac{  351264 \, \zeta_2^3 }{35} +5952 \, \zeta _5+ 38016 \, \zeta_2 \, \zeta _3  +\frac{172032}{5} \, \zeta_2^2 \,\zeta_{-1} +4096 \, \zeta_{1,-3} + 7168 \,\zeta _3 \,\zeta_{-1} \nn\\
   &+& \frac{380224 \, \zeta_2^2 }{45} +2048 \, \zeta_2  \,  \zeta_{-1}^2- 22528 \, \zeta_2 \, \zeta_{-1}-\frac{128000 \, \zeta _3}{9}-\frac{ 270848 \,\zeta_2 }{9} + \frac{55808}{27} { \Big ) } \nn\\
&+& h^{{ \bf 12 }} \,  { \Big ( } 
{ \GamOne}^{(12)}_{S=1,2} -\frac{917102848 \, \zeta_2^4 }{9975} +\frac{8888248 \, \zeta_7}{17}  + \frac{9522816 \, \zeta_2^2 \, \zeta_3}{85}-\frac{ 20301152 \, \zeta_2 \, \zeta_5}{17} -\frac{6177792}{35} \, \zeta_2^3 \, \zeta_{-1}\nn\\
   &-&  \frac{401408 \, \zeta_2 \, \zeta_{1,-3} }{3} -\frac{286720 \, \zeta_{1,-5} }{3}-\frac{705280 \, \zeta_3^2 }{3} - \frac{1735168}{3} \, \zeta_5 \, \zeta_{-1} - \frac{1495040}{3} \,\zeta_2 \, \zeta_3 \, \zeta_{-1} \nn\\
   &-&  \frac{54692992 \, \zeta_2^3 }{315} -\frac{227328}{5} \, \zeta_2^2 \, \zeta_{-1}^2 + 65536 \, ( \zeta_{1,1,-3} -  \zeta_{1,-3} \, \zeta_{-1} ) -\frac{1760896 \, \zeta_5}{9} -\frac{1138432 \, \zeta_2 \, \zeta _3 }{27} \nn\\
   &-&  57344 \zeta_3 \, \zeta_{-1}^2   - \frac{32768}{3} \, \zeta_2 \, \zeta_{-1}^3  + \frac{2136064}{15} \, \zeta_2^2 \,\zeta_{-1} +  212992 \, \zeta_{1,-3} + 372736 \zeta _3 \,\zeta_{-1}  +\frac{8532992 \, \zeta_2^2 }{45} \nn\\
&+&  106496  \, \zeta_2 \, \zeta_{-1}^2 + 278528 \, \zeta_2 \,\zeta_{-1} +\frac{15569920 \, \zeta _3}{27}   +\frac{11159552\, \zeta_2 }{27} -\frac{54956032}{243}
   { \Big )}   +   \text{O}(h^{{ \bf 14 }} ) , \nn
\eeqa
   } 
where for brevity we have grouped together a number of terms that are common to the $S$=1 and $S$=2 cases, at orders $h^{8}$, $h^{10}$ and $h^{12}$:
{\small 
\begin{align}
&\GamOne^{(8)}_{S=1,2} = -\frac{24672 \,\zeta_2^3 }{35} -3520 \, \zeta_5 -1920 \, \zeta_2^2 \, \zeta_{-1} + 864 \, \zeta_2  \, \zeta_3  -3072 \, \zeta_{1,-3} - 5376 \zeta_3 \, \zeta_{-1} - 1536 \, \zeta_2 \, \zeta_{-1}^2 ,  \nn\\
&\GamOne^{(10)}_{S=1,2} =  \frac{911816 \, \zeta_2^4 }{175}+34860 \, \zeta_7 +\frac{78336}{5} \, \zeta_2^3 \, \zeta_{-1}- 7120 \, \zeta_2  \, \zeta_5  -128 \, \zeta_2^2  \, \zeta_3 +10240 \,\zeta_{1,-5} \nn\\
&- 1024 \, \zeta_2 \, \zeta_{1,-3} - 992 \,\zeta _3^2 +19840 \, \zeta_5 \, \zeta_{-1} + 32768 \, \zeta_2  \, \zeta _3 \, \zeta_{-1} +19968 \,\zeta_2^2 \, \zeta_{-1}^2  +21504 \, \zeta _3 \, \zeta_{-1} ^2  \nn\\
&- 24576 \,(\zeta_{1,1,-3} - \zeta_{1,-3} \,  \zeta_{-1} )+ 4096 \, \zeta_2 \, \zeta_{-1}^3, \nn\\
&\GamOne^{(12)}_{S=1,2} = -\frac{11002416 \, \zeta_2^5 }{275} -335664  \, \zeta_9-\frac{125088 \, \zeta_2^3 \, \zeta_3} {35}-\frac{ 80792 \,  \zeta_2^2 \, \zeta _5}{5}+ 67340 \, \zeta_2 \, \zeta _7 - \frac{918336}{7} \, \zeta_2^4 \, \zeta_{-1} \nn \\
&- \frac{12707840 \, {\zeta_{1,-7}}}{57} 
+ \frac{29696 \, \zeta_2^2  \, {\zeta_{1,-3}}}{5} +  10240 \, \zeta_2  \, {\zeta_{1,-5}} +\frac{3358720 \, \zeta_{3,-5} }{171} -\frac{2099680 \zeta_3 \, \zeta_5 }{19} \nn\\
&+ 124672 \, \zeta_2 \, \zeta_3^2 +  142240 \zeta_7 \, \zeta_{-1} -  496640 \, \zeta_2 \, \zeta_5 \, \zeta_{-1} - \frac{620032}{5} \, \zeta_2^2 \, \zeta _3 \,\zeta_{-1} -\frac{947712}{5} \, \zeta_2^3 \, \zeta_{-1}^2 \nn\\
&-  \frac{1589248 \,\zeta_{1,3,-3}}{17} -235520 \, \zeta_{1,-3} \, \zeta _3 - 565248 \,  \zeta_2  \, \zeta_{1,1,-3} +245760 \, \zeta_{1,-5} \, \zeta_{-1}+ \frac{5357568 \, \zeta_{1,1,-5} }{17} \nn\\
&+  238080\, \zeta _5 \, \zeta_{-1}^2 -688128 \, \zeta_2 \,\zeta _3 \, \zeta_{-1}^2 - 636928 \, \zeta _3^2 \, \zeta_{-1} - 135168 \, \zeta_2^2 \, \zeta_{-1}^3  -98304 \, \zeta_{-1}^2 \, \zeta_{1, -3} \nn\\
&+ 196608 \, (\zeta_{-1} \, \zeta_{1,1,-3} - \zeta_{1,1,1,-3} ) -8192 \, \zeta_{-1} ^4 \, \zeta_2 -57344 \, \zeta_{-1}^3 \, \zeta_3 \nn .
\end{align}
}
We shall give the following results up to double wrapping. For $L$=1 and $S$=3, $Q = u^3 -\frac{5}{4} u $:
{ \small
\begin{align}
&\gamma^{L=1, S=3}= h^{\bf 2 } \, \frac{32 }{3}  - h^{\bf 4 } \, \left(\frac{224}{45}+\frac{64 \,\zeta_2 }{3}\right)\nn\\
&+ h^{ \bf 6} \,  \left( \frac{ 672 \, \zeta_2^2 }{5}  + 512 \, \zeta_3 +\frac{2048}{3} \, \zeta_2  \, \zeta_{-1}  +\frac{48448 \,  \zeta_2 }{135} -\frac{590944}{1215} \right)\nn\\
&+ h^{ \bf 8} \, { \Big (}  \GamOne^{(8)}_{S=3,4}-\frac{2647232 \, \zeta_2^2 }{675} -\frac{2658304 \, \zeta_3 }{405}  + \frac{45056}{135} \,\zeta_2  \, \zeta_{-1}  + \frac{10672576 \, \zeta_2 }{6075}+\frac{1942407808}{273375} { \Big ) } + \text{O}( h^{\bf 10}  ) , \nn
\end{align}
}
and for $L$=1 and $S$=4, $Q = u^4 -\frac{7}{2} u^2 + \frac{9}{16} $:
{ \small
\begin{align}
&\gamma^{L=1, S=4} = h^{\bf 2 } \, \frac{32
   }{3}  - h^{\bf 4 } \, \left(\frac{3872}{315} +\frac{64 \, \zeta_2 }{3}\right)\nn\\
   &+h^{ \bf 6} \,  { \Big ( } \frac{672 \,  \zeta_2^2 }{5} + 512  \, \zeta_3  + \frac{2048}{3} \, \zeta_2 \, \zeta_{-1} +\frac{151744 \, \zeta_2 }{945} - \frac{1033504}{8505} { \Big ) } +h^{ \bf 8} \, { \Big (}  \GamOne^{(8)}_{S=3,4} -\frac{13464896}{4725}  \, \zeta_2^2 \nn\\
   &\hspace{2cm} - \frac{7831552 \, \zeta_3}{2835} + \frac{2428928 \, \zeta_2 \, \zeta_{-1}}{945}
+\frac{781130944 \, \zeta_2 }{297675} + \frac{47947882624}{93767625}{ \Big ) } + \text{O}( h^{\bf 10} ) , \nn
\end{align}
}
where we have used the shorthand
{ \small 
\beqa
\GamOne^{(8)}_{S=3,4} &=& -\frac{ 32896 \, \zeta_2^3 }{35} -\frac{56320 \, \zeta_5 }{9}+ 1536 \, \zeta_2 \, \zeta_3  - \frac{10240}{3} \, \zeta_2^2 \, \zeta_{-1} -\frac{65536}{9} \, \zeta_{1,-3} \nn\\
&-&\frac{114688}{9} \, \zeta_3 \, \zeta_{-1} -\frac{32768}{9} \, \zeta_2 \,\zeta_{-1}^2 .\nn
\eeqa
}
Up to the loop order we have reached, these dimensions evaluate numerically to 
\beqa
 \gamma^{L=1,S=1} &\simeq& 8 \, h^{\bf 2 } -26.3189 \, h^{\bf 4 } + 195.487 \,  h^{ \bf 6} -1966.44\,  h^{ \bf 8} + 22309.5 \, h^{ \bf 10} -271422. \, h^{\bf 12 }  ,\nn\\
 \gamma^{L=1,S=2} &\simeq& 8 \, h^{\bf 2 } -36.9856 \, h^{\bf 4 } + 271.234 \,  h^{ \bf 6} - 2562.52\,  h^{ \bf 8}+ 28298.5 \, h^{ \bf 10} -  340713. \, h^{\bf 12 }  , \nn\\
\gamma^{L=1,S=3} &\simeq& \frac{32}{2} \, h^{\bf 2 } -40.0697 \, h^{\bf 4 } + 304.700 \,  h^{ \bf 6} -3019.43\,  h^{ \bf 8} + 33855.8 \, h^{ \bf 10} - 408283. \, h^{ \bf 12}  ,\nn\\
 \gamma^{L=1,S=4} &\simeq& \frac{32}{2} \, h^{\bf 2 } - 47.3840 \, h^{\bf 4 } + 343.369 \,  h^{ \bf 6} -3266.52\,  h^{ \bf 8} + 36262.1 \, h^{ \bf 10} - 436578. \, h^{ \bf 12} .\nn
\eeqa
More precise numerical values can be found in the attached notebook  { \tt Results.nb}.

\subsection*{Twist-2}
For the simplest twist-2 operator with $L$=2 and $S$=2, $Q = u^2 -\frac{3}{4} $, $\sigma = +1$:
{ \small
\begin{align}
&\gamma^{L=2, S=2} = 4\, h^{\bf 2 } -8 \, h^{\bf 4 } - h^{ \bf 6} \,  \left( \frac{ 84 \, \zeta_2^2 }{5}-\frac{64 \, \zeta_2 }{3} -\frac{80}{3}\right)\nn\\
&+h^{ \bf 8} \, { \Big (} \frac{1240 \, \zeta_2^3 }{7}  +300 \, \zeta_5  - 200\,\zeta_2  \,\zeta_3 -\frac{812 \, \zeta_2^2 }{5} - 192  \, \zeta_3  - 256 \, \zeta_2  \, \zeta_{-1}-64 \, \zeta_2  -112 { \Big ) } \nn\\
  &+ h^{\bf 10} \,  { \Big (} -1524 \, \zeta_2^4   -6615 \, \zeta_7 +  \frac{3264 \, \zeta_2^2 \,\zeta_3 }{5}+  3620 \, \zeta_2 \, \zeta_5  - 384 \, \zeta_2  \, \zeta_{1,-3} + 716 \, \zeta_3^2+\frac{92636 \, \zeta_2^3 }{105} \nn\\
  &+ 3100 \, \zeta_5  + 768 \, \zeta_2^2  \,\zeta_{-1} -\frac{3704}{3} \,  \zeta_2  \, \zeta_3 +
1024 \,\zeta_{1,-3} +1792 \, \zeta_3 \, \zeta_{-1} +\frac{2528 \, \zeta_2^2 }{3} + 512 \, \zeta_2 \,\zeta_{-1}^2 \nn\\
&+ \frac{3008 \,\zeta_3}{3} + 1152 \, \zeta_2 \, \zeta_{-1} + \frac{736 \, \zeta_2 }{3} + \frac{1552}{3}
   { \Big ) } \nn\\
   &+ h^{\bf 12} \,  { \Big (} 
\frac{618666 \, \zeta_2^5 }{55} +\frac{449883 \, \zeta _9}{4}
-\frac{98328}{35} \, \zeta_2^3  \, \zeta _3-\frac{54341 }{5} \,  \zeta_2^2 \, \zeta _5 -\frac{106869}{2}\, \zeta_2 \, \zeta _7  +1152\, \zeta_2^2 \, \zeta_{1,-3} \nn\\
&+ \frac{(268800 \, \zeta_{1,-7} - 17920 \, \zeta_{3,-5} )}{19} +3840 \, \zeta_2 \, \zeta_{1,-5} -\frac{577660 \, \zeta _3 \, \zeta _5}{19}+\frac{ 1880 \, \zeta_2 \, \zeta_3^2 }{3}-\frac{8516163 \,  \zeta_2^4 }{3325}\nn\\
&-  \frac{(21504 \, \zeta_{1,3,-3}-129024 \, \zeta_{1,1,-5} )}{17} -\frac{997409 \, \zeta_7}{51} + 2688 \, \zeta_{1,-3} \,  \zeta _3 + 1536 \, \zeta_2  \, \zeta_{1,1,-3} +\frac{977966 \, \zeta_2  \, \zeta _5}{51}\nn\\
&-\frac{1306172 \,  \zeta_2^2 \,\zeta_3}{255}  -\frac{9984 \, \zeta_2^3 \,\zeta_{-1} }{5}   -10240 \, \zeta_{1,-5} +6212 \, \zeta _3^2 + 3968 \, \zeta_2 \, \zeta_{1,-3} - 19840 \, \zeta _5 \,\zeta_{-1}  \nn\\
& + 7168 \, \zeta_2 \, \zeta _3 \,  \zeta_{-1} - 1536 \, \zeta_2^2 \, \zeta_{-1}^2 - \frac{1889156 \, \zeta_2^3 }{315}  -\frac{126136 \, \zeta_5}{9} +4096 \, (\zeta_{1,1,-3} -\zeta_{1,-3} \,  \zeta_{-1} ) \nn\\
&- \frac{8624 \, \zeta_2 \, \zeta _3}{9} -3584 \, \zeta _3 \, \zeta_{-1}^2 - \frac{2048}{3} \, \zeta_2 \, \zeta_{-1}^3 -\frac{171776}{15} \, \zeta_2^2 \, \zeta_{-1} - 6656 \, \zeta_{1,-3} - 4096 \, \zeta _3 \, \zeta_{-1} \nn\\
&-\frac{706352 \, \zeta_2^2 }{135} - 3328 \, \zeta_2 \, \zeta_{-1}^2   -  \frac{15584 \, \zeta_3}{3} - \frac{16640}{3} \, \zeta_2  \, \zeta_{-1} -\frac{19360 \, \zeta_2}{27} +  \frac{1552}{3} { \Big ) } + \text{O}( h^{\bf 14} ), \nn
  \end{align}
  which evaluates to
  \beqa
  \gamma^{L=2,S=2} \simeq 4 \, h^{\bf 2 } -8 \, h^{\bf 4 } + 16.301018 \,  h^{ \bf 6} + 108.45083\,  h^{ \bf 8} - 3197.3744 \, h^{ \bf 10} + 56169.236 \, h^{\bf 12 }  
  .\nn
\eeqa
}

\section{Symmetry}\label{app:symmetry}

In this Appendix we shall discuss a simple symmetry of the QSC equations, which allows us to choose freely the constants $A_1$ and $A_2$ and four of the coefficients of the polynomials (\ref{eq:polyMK}). 
 The simplest symmetry of the QSC equations (see \cite{Gromov:2014caa} and Section 2.3 of \cite{Gromov:2014bva} for the very similar $\mathcal{N}$=4 case) is the transformation 
\beqa\label{eq:nuR}
\nu_a \rightarrow R_{a}^{ \; b} \, \nu_b , \;\;\;\;\;{ \bP }_{ab} \rightarrow R_{a}^{ \; i} \, { \bP }_{ij} \, R_{b}^{ \; j} ,
\eeqa
where $R$ is any $4 \times 4$ constant matrix satisfying
\beqa\label{eq:Rconstr}
R^{t} \, \chi \, R = \chi .
\eeqa
To preserve the structure of the QSC, we should also require that the transformation does not change the ordering of the magnitudes of the $\bP$ functions at large $u$, $ |{\bP}_2 |< |\bP_1 |< |\bP_0 |< |\bP_4| < |\bP_3|$. 
The most general form of $R$ compatible with these constraints has six degrees of freedom and can be written as:
\beqa
R = \left(
\begin{array}{cccc}
 \alpha_1 \; &  0\; & 0\; & 0\; \\
 \alpha_2 \;& \alpha_3\; & \alpha_4\;  & 0\;  \\
-\frac{\alpha_5}{\alpha_3} \;& 0 \;& \frac{1}{\alpha_3}\; & 0\\
\frac{ \alpha_1  \alpha_6 + \alpha_2 \alpha_5}{\alpha_1 \alpha_3} \;& \frac{\alpha_5 }{\alpha_1} \; & \frac{\alpha_2 + \alpha_4 \alpha_5 }{\alpha_1 \alpha_3} \;& \frac{1}{\alpha_1}\; \\
\end{array}
\right) ,
\eeqa
 and the transformation acts on the $\bP$ functions as
\beqa
\bP_1 &\rightarrow& \alpha_1 \, \left( \alpha_3 \bP_1 + \alpha_4 \bP_2 \right), \;\;\;\;\; \bP_2 \rightarrow  \frac{\alpha_1}{\alpha_3} \, \bP_2 , \label{eq:symP12}\\
\bP_0 &\rightarrow& \bP_0  +\alpha_5 \, \bP_1 + \frac{\alpha_2 + \alpha_4 \alpha_5 }{\alpha_3} \, \bP_2 , \label{eq:symP0}\\
\bP_3 &\rightarrow & \frac{\alpha_3}{\alpha_1} \, \bP_3 + \frac{\alpha_4}{\alpha_1} \,\bP_4 + 2 \, \frac{\alpha_2}{\alpha_1} \, \bP_0 -\alpha_6 \, \bP_1 + \frac{ \alpha_2^2 - \alpha_1 \alpha_4 \alpha_6 }{\alpha_1 \alpha_3} \,  \bP_2 \label{eq:symP3}, \\
\bP_4 &\rightarrow & \frac{1}{\alpha_1 \, \alpha_3 } \,\left( \bP_4 - 2 \, \alpha_5 \, \bP_0 - \alpha_5^2  \, \bP_1 - ( \alpha_4 \alpha_5^2 + 2  \alpha_2 \alpha_5  + \alpha_1 \alpha_6 ) \,  \bP_2 \right). 
\eeqa
From (\ref{eq:symP12}), it is simple to see that the constants $A_1$ and $A_2$ characterising the leading asymptotics $\bP_1 \sim A_1 u^{-L}$, $\bP_2 \sim A_2 u^{-L-1}$ can be fixed to arbitrary values by an appropriate choice of $\alpha_1$ and $\alpha_3$. 
 Similarly, relations (\ref{eq:symP0}),(\ref{eq:symP3}) show that the coefficients $m_0(h)$, $k_0(h)$, $k_L(h)$ and $k_{2L}(h)$ defined in (\ref{eq:polyMK}) -- which correspond to certain coefficients of the large-$u$ expansion of $\bP_0$ and $\bP_3$ -- can be chosen freely by tuning $\alpha_5$, $\alpha_6$, $\alpha_2$ and $\alpha_4$, respectively. 
 Accordingly, these numbers are not fixed by the algorithm at any order in $h$.

\section{Solving inhomogeneous Baxter equations}
\label{app:inhomo}
In this Appendix, we shall present the basic method to solve the inhomogeneous Baxter equations (\ref{eq:inhNu1}) and (\ref{eq:inh2}) encountered in the procedure. 
\subsection{Solving the inhomogeneous Baxter equation for $\nu_1$}\label{app:inh1}

At a generic perturbative order $\text{O}(h^{2n})$, equation (\ref{eq:inhNu1}) reduces to
\beqa\label{eq:eqnu1}
  (u+i/2)^L \,  q_{1}^{[2]} -  (u - i/2)^L \, q_{1}^{[-2]}  - T_0 \, q_{1}   =  - U_1^{[-1]} ,
\eeqa
where $ q_1^{} = \nu_{1, \text{ns}, n}^{[1]}$, $U_1$ a source term of increasing complexity, and $T_0$ is the zero-order transfer matrix.  At leading order, we know that the source term is zero, and the regular solution  is the Baxter polynomial $Q$. 
To solve the generic case, we follow the method of \cite{Marboe:2014gma}.  Considering the ansatz $q_1=Q \, f_1^{[1]}$, it is simple to see that, in order for (\ref{eq:eqnu1}) to be fulfilled, $f$ must satisfy
\beq
\nabla_{-}\( u^L Q^{[1]} Q^{[-1]} \nabla_{+}(f_1)\) = U_1 \, Q^{[1]} ,
\eeq
where we have denoted 
$
\nabla_{+} \, g = g - g^{[2]} $ and $\nabla_{-} \, g = g + g^{[2]} $.  
We introduce the inverse operators $\Psi_+$ and $\Psi_-$, such that $\nabla_{\pm} \Psi_{\pm} g = g$. We shall give a precise operative definition of the operators $\Psi_{\pm}$ in Section \ref{app:PsiPsibar}, by explaining how they act on the functions generated by the algorithm. 
 Using these operators, an inhomogeneous solution of (\ref{eq:eqnu1}) can be found as   
\beq\label{eq:finh1}
f_{1, \text{inhomo}}=\Psi_+ \( \frac{1}{u^L Q^{[1]} Q^{[-1]}}  \Psi_- \(U_1 Q^{[1]} \) \) . 
\eeq
 Moreover, setting $U_1=0$ and $\Psi_-(0) = \Phi_{1, \text{anti}} $, where $\Phi_{1, \text{anti}} = - \Phi_{1, \text{anti}}^{[2]}$  is a  generic anti-symmetric function, we find a second independent solution of the homogeneous equation, 
\beq\label{eq:1homo}
 f_{1, \text{homo}} = \Phi_{1, \text{anti}}  \; \Psi_- \( \frac{1}{u^L Q^{[1]} Q^{[-1]}} \) .
 \eeq
Introducing for convenience the notation
\beqa\label{eq:defineho}
\ho^{[-1]} = Q^{[-1]} \, \Psi_- \( \frac{1}{u^L Q^{[1]} Q^{[-1]}} \) ,
\eeqa
 and putting all pieces together, the general solution of (\ref{eq:eqnu1}) can be written as 
\beq\label{eq:geninhomsol1}
q_1^{[-1]} =\Phi_{1, \text{per} } Q^{[-1]} + \Phi_{2, \text{anti}} \, \ho^{[-1]}  + \Psi_+ \( \frac{1}{u^L Q^{[1]} Q^{[-1]}}  \Psi_- \(U_1 Q^{[1]} \) \) ,
\eeq
where $\Phi_{1, \text{per}} = \Phi_{1, \text{per}}^{[2]}$ denotes a generic $i$-periodic function. 
Following \cite{Marboe:2014gma}, it is convenient to rewrite this expression in order to cancel its apparent poles at the Bethe roots where $Q=0$. To achieve this, we introduce two polynomials $A$ and $B$, of degree $S-1$, defined by the condition 
\beqa
A \, Q^{[1]} + B \, Q^{[-1]}=1 .
\eeqa 
The Baxter equation then implies that 
\beqa
-A^{[1]}(u^{[-1]})^L+B^{[-1]}(u^{[1]})^L=Q R,
\eeqa
 with $R$ a polynomial of degree $L-2$. Introducing the constants $r_{\pm, k}$ and the polynomial $C$ through
\beqa\label{eq:Crpm}
\frac{R}{(u^{[1]} u^{[-1]})^L}&=& \sum_{k=1}^{L}\( \frac{r_{k,+}}{(u^{[1]})^k}+\frac{r_{k,-}}{(u^{[-1]})^k}  \) , \\
C &=& \frac{A}{u^L}-Q^{[-1]} \, \sum_{k=1}^L \frac{ r_{k, +} }{u^k} ,
\eeqa
 we see that (\ref{eq:defineho}) can be written as
\beq \label{eq:sechomosol1}
\ho^{[-1]} = \( C + Q^{[-1]}\sum_{k=1}^{L} \(-r_{k,+}+r_{k,-}  \) \eta_{-k}(u)       \) ,
\eeq
where $\eta_{-k}(u)$, $k \in \mathbb{N}^+$ are certain polygamma functions (see equation (\ref{eq:polygamma2}) below), solutions of 
\beqa
\eta_{-k} + \eta_{-k}^{[2]} = \frac{1}{u^k} .
\eeqa
Likewise, the inhomogeneous solution of (\ref{eq:finh1}) can be written in the following form, which only involves poles at positions $u \in i \mathbb{Z}$:
\beqa\label{eq:finh1b}
f_{1, \text{inhomo}} &=& 
\Psi_- \(U_1 Q^{[1]} \)C+   Q^{[-1]} \Psi_+ \(\Psi_-  \(U_1 Q^{[1]} \) \sum_{k=1}^L \frac{-r_{k,+}+r_{k,-}}{u^k} +C^{[2]}U_1  \) . \nn \\
&&
\eeqa

\subsection{ Solving the inhomogeneous Baxter equation for $\nu_2$}\label{app:inh2}

The equation arising from (\ref{eq:inh2}) at the $(n+1)$-th iteration of the algorithm has the form
\beqa\label{eq:eqnu2}
 (u+i/2)^L \,  q_{2}^{[2]}  -  (u - i/2)^L \,  q_{2}^{[-2]}  + T_0 \, q_{2}   =  - U_2^{[-1]} ,
\eeqa 
with $ q_2^{} = \nu_{2, \text{ns}, n}^{[1]}$. One can proceed in a similar way as for (\ref{eq:eqnu1}), by paying attention to the different sign in front of the $T_0 \, q_2$ term. This implies that a simple, homogeneous solution of the equation for $U_2=0$ is simply given by $q_2 = Q \, \Phi_{2, \text{anti}}$, where $\Phi_{2,\text{anti}} = - \Phi_{2, \text{anti}}^{[2]}$ is any anti-periodic function. Likewise, we see that an independent family of solutions to the homogeneous equation is described by
\beqa
q_2 = \Phi_{2, \text{per}}^{[-1]} \,  \ho ,
\eeqa
where $\Phi_{2, \text{per}} = \Phi_{2, \text{per}}^{[2]}$ is $i$-periodic  and $\ho$ is defined in (\ref{eq:sechomosol1}).

Making the ansatz $q_2 = Q \, f_2^{[1]}$, we find 
\beq
\nabla_+ \( u^L Q^{[1]} Q^{[-1]} \nabla_- (f_2)\) = - U_2 Q^{[1]} ,
\eeq
 so that a solution to the inhomogeneous equation can be found as  $q_2 = Q \, f_{2, \text{inhomo}}^{[1]}$, with
\beq\label{eq:f2inh}
f_{2, \text{inhomo}} = \Psi_-  \( \frac{1}{u^L Q^{[1]} Q^{[-1]}}  \Psi_+ \(-U_2 Q^{[1]} \) \) ,   
\eeq
and the most general solution can be written as 
\beq\label{eq:geninhomsol2}
q_2^{[-1]} =\Phi_{2, \text{anti}} Q^{[-1]} + \Phi_{2, \text{per}} \ho  + \Psi_+ \( \frac{1}{u^L Q^{[1]} Q^{[-1]}}  \Psi_- \(U_1 Q^{[1]} \) \) .
\eeq
More explicitly, in terms of the same polynomial $C$ and constants $r_{k, \pm}$ defined in (\ref{eq:Crpm}), (\ref{eq:f2inh}) can be expressed as
\beqa \label{eq:inhomosol2}
&&Q^{[-1]} \,   f_{2, \text{inhomo}} \\
&=&\Psi_+ \(-U_2 Q^{[1]} \)C+Q^{[-1]} \Psi_- \( \Psi_+ \(-U_2 Q^{[1]} \) \sum_{k=1}^L \frac{-r_{k,+}+r_{k,-}}{u^k} - C^{[2]} U_2  \).\nn
\eeqa

\subsection{Periodic coefficient functions}\label{sec:perPhi}
 To construct periodic/anti-periodic functions without introducing unphysical poles or an unphysical exponential growth at infinity, we consider the following combinations
 \beqa\label{eq:Peri}
\mathcal{P}_k(u) = \eta_k(u) + \text{sgn}(k) \, { \bar \eta}_{k}^{[-2]}(u) =\text{sgn}(k) \, \mathcal{P}_k(u + i) , \;\;\;\; 0 \neq k \in \mathbb{Z} ,
\eeqa
where $\eta_k$ is defined in (\ref{eq:polygamma}). The coefficient functions appearing in  (\ref{eq:geninhomsol1}), (\ref{eq:geninhomsol2}) are then constructed, at every iteration of the program, as
\beqa\label{eq:percoeffs}
\Phi_{a, \text{per} }(u) = \phi_{a, 0}^{\text{per}} +  \sum_{j=1}^{\Lambda} \phi^{\text{per}}_{a, j} \, \mathcal{P}_j(u) , \;\;\;\;\;\; \Phi_{a, \text{anti} }(u) = \sum_{j=1}^{\Lambda} \phi^{\text{anti}}_{a, j} \, \mathcal{P}_{-j}(u) ,  
\eeqa
for $a=1,2$, where $\phi^{\text{per}}_{a, j}$, $\phi^{\text{anti}}_{a, j} $ are free parameters. 
 The number of terms included in the sum increases with the perturbative order: at the $n$-th iteration, we may take $\Lambda = 2(n - 1)$.

\section{ Functions generated by the algorithm }\label{app:functions}

In this Appendix, we give a precise definition of the operations $\Psi_{\pm}$. This will show that the algorithm always produces answers in a specific algebra of functions comprising: 
\begin{itemize}
 \item 
rational functions of $u$, with poles only in the set $u \in i \mathbb{Z}$ ;
\item  the functions $\eta_{A}$, with $A$ a multi-index $A = a_1, a_2, \dots , a_k$, $a_i \in \mathbb{Z} \setminus \left\{0\right\}$, which are a generalisation of Hurwitz zeta functions \cite{leurent2013multiple} , 
\item
periodic/anti-periodic functions constructed as explained in Section \ref{sec:perPhi}.
\end{itemize}

The  presence of alternating signs is an important difference as compared to the ${\cal N}$=4 SYM case, leading to the appearance of  alternating Euler-Zagier sums in the results. 

\subsection{ The $\eta$ functions }

When the sums are convergent, the operators $\Psi_{+}$ and $\Psi_{-}$ can be implemented as 
\beq
\Psi_+( g )=\sum_{n=0}^{\infty}\ g^{[+2n]}  ,   \qquad    \Psi_-(g)=\sum_{n=0}^{\infty}(-1)^n\ g^{[+2n]}  .
\eeq
This leads to the natural definition of the $\eta_a$ functions satisfying $\nabla_{\pm} \eta_{\pm|a|}(u) = \frac{1}{u^{|a|} }$: 
\beqa\label{eq:etadef1}
\eta_a(u) = \sum_{n=0}^{\infty} \frac{(\text{sgn}(a))^n }{ (u+i n)^{|a|} } ,
\eeqa
which is simply expressible in terms of polygammas:
\beqa
\eta_{a}(u) &=& \frac{i^a }{(a-1)!}\, \psi^{(a-1)}(-i u) , \;\;\;\; a \in \mathbb{N}^+ , \label{eq:polygamma}\\
\eta_{-a}(u) &=& \frac{i^a }{ 2^a \, (a-1)!}\, \left[ \psi^{(a-1)}\left(-\frac{iu}{2}\right) -  \psi^{(a-1)}\left(\frac{1}{2} -\frac{iu}{2} \right) \right] , \;\;\;\;  a \in \mathbb{N}^+  .\label{eq:polygamma2}
\eeqa
It should be noted that the sum (\ref{eq:etadef1}) is divergent for $a=1$, but we may use (\ref{eq:polygamma}) as regularised definition of $\eta_1$.  The ambiguity in this choice does not affect any physical results. In fact,  notice that the $\eta$ functions enter the algorithm through the solution of inhomogeneous Baxter equations and this ambiguity only goes into a redefinition of the integration constants described in Section \ref{app:inhomo}. Another  important point to underline is that the particular regularisation of $\eta_1$ introduces $i \eta_1(i)\equiv  \zeta_1^{\text{reg}} \equiv \gamma_{\text{Euler-Mascheroni}}$ in the solution of the QSC. While this number enters the expansions of various $\bP$ and $\nu$ functions (which are not directly physical due to the symmetry described in Appendix \ref{app:symmetry}), it always cancels out of the anomalous dimensions. 

We then define $\eta_{a, B}$, with $B$ any multi-index, as follows
\beqa\label{eq:etadef2}
\eta_{a, B}(u) = \sum_{n=0}^{\infty} (\text{sgn}(a))^n \frac{\eta_{B}^{[2 n+2]}(u) }{(u+i n )^{|a|} } .
\eeqa
Explicitly, these functions can be written as
\beqa\label{eq:etadef} 
\eta_{a_1, a_2, \dots , a_k }( u ) = \sum_{ n_k > n_{k-1} > \dots > n_1 \geq 0} \prod_{i=1}^k \frac{(\text{sgn}(a_i))^{n_i - n_{i-1} - 1 } }{( u + i n_i )^{|a_i|} },   
\eeqa
where by convention we set $n_0 = -1$. The sum is convergent for $a_k \neq 1$. 
 One can define the marginally divergent cases where $a_k = 1$ through the \emph{stuffle algebra} (see Appendix A of \cite{leurent2013multiple}), which allows one to express them in terms of convergent $\eta$ functions and $\eta_1$. 

 Relation (\ref{eq:etadef}), shows that the Laurent expansion of the $\eta$ functions around $u=i$ naturally produces multiple Euler-Zagier sums, which therefore enter the algorithm. The precise relation is 
 \beqa
 \eta_{A}(i) = i^{-|A| } \,N_A \, \zeta_{\bar{A}} ,
 \eeqa
 where, given the multi-index $A = \left\{ a_1, \dots, a_k \right\}$, we have defined 
 \beqa
 |A| = \sum_{j=1}^{k} |a_k|, \;\;\;\;\;
 N_A = \prod_{j=1}^k \text{sgn}( a_j ) , 
 \eeqa
 and the multi-index $\bar{A}$ is 
 \beqa
 \bar{A} &=& \left\{ a_1 \, \text{sgn}(a_2) ,  \dots , a_m \, \text{sgn}( a_{m+1} ) , \dots , a_k \right\}.
 \eeqa  
Alternatively, it is useful to note the following relation with the harmonic polylogarithms $H_{A}(x)$ of \cite{Remiddi:1999ew} evaluated at unity (see also Section 10 of \cite{Maitre:2005uu}):
\beqa\label{eq:relationHarm}
\eta_{A}(i) = i^{-|A| } \, H_{ { \widetilde A } }(1) ,
\eeqa
where ${ \widetilde A }$ is obtained by reversing the order of the indices in $A$. Relation (\ref{eq:relationHarm}) is the most useful for comparisons with the notation of the datamine \cite{DataMine}.

\subsection{Defining the operators $\Psi_{\pm}$ }\label{app:PsiPsibar}

 Let us finally come to the full definition of the linear operators $\Psi_{\pm}$. Applied to a polynomial, we require that they  yield a polynomial answer of the form
 \beqa
 \Psi_-\left(  \sum_{j=0}^m \, a_j \, u^j \right) = \sum_{j=0}^m \, b_j \, u^j  ,\;\;\;\;\; \Psi_+\left( \sum_{j=0}^m a_j \, u^j  \right) = \sum_{j=1}^{m+1} c_j \, u^j . 
 \eeqa
 Each rational function of $u$ is then broken into a polynomial part and a sum of inverse powers, leading to the appearance of $\eta$ functions
 \beqa
 \Psi_{\pm}\left( \frac{1}{u^{|a|} }\right) = \eta_{\pm |a|}.
 \eeqa
 As in \cite{Marboe:2014gma}, we then define recursively the action of $\Psi_{\pm}$ on the products of rational functions and $\eta$'s. One starts form noticing the following relation (which is a consequence of the nested definition (\ref{eq:etadef2})) 
 \beqa\label{eq:trasl}
\eta_{a, A} = ( \text{sgn}(a) ) \, \eta_{a, A}^{[2]} + \frac{\eta_{A}^{[2]}}{u^{|a|}} ,
\eeqa
for any multi-index $A$, which leads immediately to 
\beqa\label{eq:nice}
\Psi_{\pm}\left( \frac{\eta_{A}^{[2+2n]} }{(u+i \, n)^{|a|} } \right) = \eta_{ \pm |a|, A}^{[2 n]} .
\eeqa
Besides, when more general combinations like $\eta_{B}^{[2m]}/(u+i \, k)^{|a|}$ are encountered, one can use (\ref{eq:trasl})  until either a term of the form (\ref{eq:nice}) is met, or the $\eta$ runs out of indices ($\eta_{\emptyset}=1$). 
To deal with the polynomial-times-$\eta$ part, we use the relation
\beqa
\nabla_{\pm}\left( \Psi_{ \pm \text{sgn}(b) }(u^{|a|}) \, \eta_{b , A}^{[2n]} \right) &=& u^{|a|} \eta_{b, A}^{[2n]} \pm (  \text{sgn}(b) ) \,  \Psi_{ \pm \text{ign}(b) }(u^{|a|})^{[2]} \, \frac{\eta_{A}^{[2 + 2 n]}}{(u+ i n )^{|b|} } , 
\eeqa
which can be checked by a little algebra, and finally obtain
\beqa
\Psi_{\pm } \left( u^{|a|} \, \eta_{b, A}^{[2n]} \right) &=& \Psi_{  \pm \text{sgn}(b) }( u^{|a|} ) \, \eta_{b, A}^{[2n]} \mp (  \text{sgn}(b) ) \Psi_{\pm} \left(  \Psi_{  \pm \text{sgn}(b) }(u^{|a|})^{[2]} \, \frac{\eta_{A}^{[2 + 2 n]}}{(u+ i n )^{|b|} } \right) .\nn\\
\eeqa
 The $i$-periodic/antiperiodic functions $\mathcal{P}_k$ are easily dealt with using
 \beqa
 \Psi_{\pm}\left( \mathcal{P}_k \, \dots \right) = \mathcal{P}_{k} \, \Psi_{\pm (\text{sgn(k)})}\left(  \dots \right) .
\eeqa
In conclusion, the operators $\Psi_{\pm}$ are fully defined on the algebra of trilinear combinations of rational, $\eta$ and $\mathcal{P}_k$ functions. We notice that, since some of the equations adopted in the algorithm are quadratic, products of two $\eta$ functions may also be generated along the way. However, by using stuffle algebra relations such as the ones in \cite{leurent2013multiple}, these can be converted into expressions that are linear in all the $\eta$ functions. Up to the order we reached, it actually turned out that this step was unnecessary, since all the source terms in the inhomogeneous Baxter equations were already linear in the $\eta$'s.

\bibliography{biblio7} 

\end{document}